\documentstyle[twocolumn,aps,psfig]{revtex}

\newcommand\beq{\begin{equation}}
\newcommand\eeq{\end{equation}}
\newcommand\bea{\begin{eqnarray}}
\newcommand\eea{\end{eqnarray}}

\begin{document}

\title{The Mixmaster Model as a Cosmological Framework
and Aspects of its Quantum Dynamics}

\author{Giovanni Imponente$~^{1,2}$ {\thanks{E-mail address: 
imponente@icra.it}} and
Giovanni Montani$~^{1,3}$ \thanks{
E-mail address: montani@icra.it} }
\address{
$~^1$ICRA---International Center for Relativistic Astrophysics, Piazza A.Moro 5 00185 Rome, Italy \\  
$~^2$Universit\`a degli Studi di Napoli ``Federico II'' and INFN---Sezione di Napoli \\
$~^3$Physics Department, University of Rome ``La Sapienza''} \maketitle

\begin{abstract}
{\bf {Abstract:}} 
This paper provides a review of some recent issues 
on the Mixmaster dynamics concerning
the features of its stochasticity. 
After a description of the geometrical structure 
characterizing the homogeneous cosmological models 
in the Bianchi classification and the 
Belinsky-Khalatnikov-Lifshitz piecewise representation 
of the types VIII and IX oscillatory regime, we face 
the question regarding the time covariance of the resulting
chaos as viewed in terms of continuous Misner-Chitr\'e like
variables. Finally we show how in the statistical mechanics 
framework the Mixmaster chaos raises as semiclassical
limit of the quantum dynamics in the Planckian era.

\vspace{0.5cm}

PACS number(s): {04.20.Jb, 98.80.Dr}
\end{abstract}
\smallskip\

\section{Introduction}

As well-known the Standard Cosmological Model (SCM)
finds its theoretical basis 
in terms of a homogeneous and isotropic universe obtained as high 
symmetry solution of the Einstein's equations, the so called 
Friedmann-Robertson-Walker model (FRW). 
Such representation of our 
actual universe possesses a clear degree of reliability due to its 
good general agreement with respect to the observed phenomenology 
(in particular the strong isotropy 
of the Cosmic Background Radiation as well as the consistency between the 
predictions on the primordial nucleosynthesis of the light elements and 
the experimentally observed abundances), nevertheless there are some important 
general aspects to be taken into account. 

In first place none theoretical principle led us to exclude that in the 
very early phases of its evolution the universe had been characterized by 
a higher degree of inhomogeneity and anisotropy and only in a later stage 
underwent an isotropization process as natural consequence of its dynamics 
and/or by the action of some physical mechanism in classical as well as 
quantum regime 
(to which the causality notion plays a crucial role);
indeed the instability of the FRW solution toward the cosmological 
singularity implies that a more general behavior should characterize
the very early evolution as soon as we make allowance 
for small perturbations responsible for the actual
clumpyness. 

Furthermore the SCM in its original formulation 
contains non-trivial internal inconsistencies which require 
to be explained through appropriate modifications of the underlying theory. 
Among this undesired theoretical facts take particular importance the 
so called ``horizon paradox'' and ``flatness problem'' to which should be added 
the absence of an appropriate model for the large-scale structures 
formation able to reproduce the observed distribution of matter in the 
actual universe \cite{KO1,KO2,KO3}. 

All  these considerations  make clear the 
deep interest in studying more general 
classes of solutions of the Einstein's equations in order to individualize 
dynamical behaviors which could constitute a more suitable framework than 
the simple FRW model for the construction of a 
completely self-consistent cosmology, with particular reference to the 
universe evolution in proximity of the initial ``Big Bang'' \cite{KO4,KO5,KO6}. 

In this paper (Sections I and II) we will discuss one of the most studied cosmological 
dynamical models in view of its possible implications in the history of our 
universe, the so called Mixmaster model \cite{M69}\cite{BKL70,BKL82}. 
It corresponds to the asymptotic 
evolution toward the cosmological singularity of the type models VIII and IX 
in the famous Bianchi classification \cite{LF}. Another important theoretical 
property characterizing such model is that it is an appropriate prototype 
of the behavior of the general cosmological solution 
of the Einstein's equations in the same 
asymptotic region to the initial singularity. 
Indeed the direct extension to the inhomogeneous case of the 
ideas and the formalism required by the treatment of the homogeneous one, 
extension based on the dynamical 
decoupling characterizing near the singularity the different points of the 
space, leads to derive in natural way the asymptotic evolution 
to the ``Big Bang'' of a generic  
inhomogeneous cosmological model which constitutes surely one of the most 
important results until now obtained in Relativistic Cosmology 
\cite{BKL70,BKL82}.  

The description of the Mixmaster Model can take place in terms of two 
different, but indeed completely equivalent approaches: one is based on the 
direct analysis of the Einstein's equations, as in the 
original analysis due to Belinski, Khalatnikov and 
Lifshitz (BKL) (Section III), while the 
other corresponds to a Hamiltonian approach to the dynamics (Section IV)
which had been introduced first by Misner \cite{M69}.

However not all 
anisotropic dynamics are compatible with a satisfactory SCM but, as shown 
in the early Seventies, under suitable conditions some can be represented  
as a FRW model plus a gravitational waves packet (\cite{L74}, \cite{GDY75}). \\
Among the Bianchi classification, the types VIII\footnote{All the considerations we will 
develop for the type IX apply also to the VIII one since, 
close to the singularity, they have the same morphology.} 
and IX appear as the most general ones: 
the former's geometry is 
invariant under the $SO(3)$ group, 
like the closed FRW universe 
and its dynamics allows the line element 
to be decomposed as
\begin{equation}
\label{ds}
ds^2={ds_0}^2 -\delta_{(a)(b)}G^{(a)(b)}_{ik}dx^i dx^k
\end{equation}
where $ds_0$ denotes the line element of an isotropic universe having 
positive constant curvature, $G^{(a)(b)}_{ik}$ is a 
set of spatial tensors\footnote{These tensors satisfy the equations 
\[
G^{(a)(b);l}_{~ik~~;l}= -(n^2-3)G^{(a)(b)}_{~ik}, \quad 
G^{(a)(b)k}_{~~;k}=0, \quad
G^{(a)(b)i}_{~~i}=0 \, , 
\]
in which the Laplacian is referred to the geometry of the sphere of unit radius.}
and $\delta_{(a)(b)}(t)$ are amplitude functions, 
resulting small sufficiently far
from the singularity. \\
Since Belinski, Khalatnikov and Lifshitz \cite{BKL70}
derived the oscillatory regime characterizing 
the evolution near a physical singularity, a wide literature 
faced over the years this subject in order to understand 
the resulting chaos. 

The research developed overall in two different, 
but related, directions: firstly, the dynamical analysis 
removed the limits of the BKL approach due
to its discrete nature, providing satisfactory representations
in terms of continuous variables (i.e. construction 
of an invariant measure for the system \cite{CB83}, \cite{KM97}); 
secondly, a better characterization of 
the Mixmaster chaos (in view of its properties of covariance)
found non-trivial difficulties  
to apply the standard chaos indicators to relativistic systems. \\
The various approaches prevented, up to now, to say a definitive word
about the covariance of Mixmaster chaos, though important 
indications in favor of such covariance arise 
from \cite{CL97}, \cite{ML00}, \cite{IM01} and \cite{IM01a,IM01b}, 
as above discussed in detail.

A wide interest in understanding 
this intrinsic nature of the chaotic Mixmaster 
dynamics and its very early appearance in the Universe
evolution, to which we dedicate 
our analysis, 
lead to believe in the existence of a relation
with the quantum behavior the system performs during the 
Planckian era.
In fact the last aim of this paper (Section V) 
is to give a precise meaning to such 
relation by constructing the semiclassical limit of a Sch\"oedinger
approach to the canonical quantization of the 
Arnowitt-Deser-Misner (ADM) dynamics.

\section{Geometrical Structure of the Bianchi Models}

\subsection{Group Transformations}

A (pseudo) Riemannian space, given a metric $g$ 
and a differential 
structure $(M,g)$,
is homogeneous if it is orbit for an isometry 
group as a group invariance
for the metric $g$. The groups of movements for the 
metric are said to be isometry groups and for homogeneous spaces 
they are transitive, in the sense that each point is equivalent 
under the action of the group and they are spatial sections of 
the space-time. \\
We are interested in the Lie algebras for Killing vector fields
as generators of these groups of movements, intended as 
infinitesimal transformations generating rotations and translations. \\
Let's consider a group of transformations
\beq
\label{e1}
x^{\mu} \rightarrow \bar x^{\mu} = f^{\mu}\left(x,a\right)
\eeq
over a space $M$, where $\left\{ a^a \right\}_{a=1,..,r}$ are 
$r$ independent variables parameterizing the group and  
$a_0$ corresponds to the identical transformation 
\beq
\label{e2}
f^{\mu}\left(x,a_0 \right) = x^{\mu}\, .
\eeq
Let's then consider also the infinitesimal transformation
corresponding to $a_0+\delta a$, close to the identity,
\bea
\label{e3}
x^{\mu} \rightarrow \bar x^{\mu} 
&=&f^{\mu}\left(x, a_0+\delta a \right) \approx \nonumber \\
&\approx& \underbrace{ f^{\mu}\left(x, a_0 \right)}_{=x^{\mu}} + \underbrace{\left( \frac{\partial f^{\mu}}{\partial a^a} \right) \left(x, a_0 \right) }_{\equiv\xi^{\mu}_a\left(x\right)}\delta a^a
\eea
say
\beq
\label{e4}
x^{\mu} \rightarrow \bar x^{\mu} 
\approx x^{\mu} + \xi^{\mu}_a\left(x\right)\delta a^a
=\left(1+\delta a^a \xi_a \right) x^{\mu} \, ,
\eeq
where the $r$ first order differential operators 
$\left\{\xi_a\right\}$ are defined as 
$\xi_a=\xi^{\mu}_a\frac{\partial}{\partial x^{\mu}}$ 
in correspondence with the $r$ vectorial fields 
with components $\left\{ \xi^{\mu}_a\right\}$.
These are the Killing generating vector fields.\\
The coordinate transformation can be definitely 
written as 
\beq
\label{e5}
\bar x^{\mu} \approx \left( 1+ \delta a^a \xi_a \right) 
\approx e^{\delta a^a \xi_a} x^{\mu} \, 
\eeq
and in a finite form 
\beq
\label{e6}
\bar x^{\mu} \rightarrow \bar x^{\mu} =e^{\theta^a \xi_a} x^{\mu}
\eeq
where $\left\{ \theta^a \right\}$ are $r$ new 
parameters for the group. \\
The vectorial generator fields form a Lie algebra,
say a real r-dimensional vectorial space with 
basis $\left\{\xi_a\right\}$,
closed with respect to commutation, in order to have 
the representation 
\beq
\label{e7}
\left[ \xi_a ,\xi_b \right] \equiv \xi_a \xi_b - \xi_b \xi_a 
= C^c_{~ab} \xi_c
\eeq 
where $C^c_{~ab} $ are the structure constants for the 
Lie algebra. \\
It is then natural to extend this formalism
to the choice of $\left\{ e_a \right\}$ as basis 
for the Lie algebra $g$ over a group $G$ with 
\beq
\label{e8}
\left[ e_a , e_b \right] = C^c_{~ab} e_c \, ;
\eeq
then let's define the symmetric quantity
\beq
\label{e9}
\gamma_{ab} =C^c_{~ad} C^d_{~bc} =\gamma_{ba}
\eeq
and an internal product over the Lie algebra 
\beq
\label{e10}
\gamma_{ab} \equiv e_a \cdot e_b 
= \gamma\left( e_a , e_b \right) \, , 
\quad \gamma \left(X^a e_a, Y^b e_b \right) 
=\gamma_{ab} X^a Y^b \, .
\eeq
The orbit of $x$ is 
\beq	
\label{e17}
f_G \left( x \right) 
= \left\{ f_a \left( x \right) \mid a \in G \right\} 
\eeq
as the set of all points that can be 
achieved from $x$ under the group of 
transformations. 
The isotropy group in $x$ is 
\beq
\label{e18}
G_x = \left\{ a \in G \mid f_a \left(x \right) =x \right\} \, ,
\eeq
i.e. the subgroup of $G$ which leaves $x$ fixed.
If $x_1,x_2$ are on the same orbit, then 
$G_{x_1}$ and $G_{x_2}$ are conjugate subgroups of $G$
and then isomorphic.
If now $G_x=\left\{a_0\right\}$ and 
$f_G\left(x\right)=M$, $G$ is diffeomorphic to $M$ and 
the two spaces are identified. If $g$ is 
a metric over $M$ invariant under $G$, then
it is definitively given by the 
inner product of the invariant vectorial fields
of basis $e_a$. 

The groups of non-Abelian transformations, as defined by 
(\ref{e8}), represent geometrically homogeneous 
spaces in three dimensions and give the section spatially
homogeneous of the spatially homogeneous space-times. \\
Given a basis $\left\{e_a \right\}$ of the 
Lie algebra for the tridimensional Lie group, 
with structure constants $C^c_{~ab}$, 
at any time the spatial metric is given by 
the spatially constant inner products
\beq
\label{e19}
e_a \cdot e_b =g_{ab} \left(t \right) \, \qquad a,b=1,\ldots,3
\eeq
which are six functions of the time variable only.
This permits to rewrite the Einstein equations 
as ordinary differential equations, 
eventually associated to constraints functions 
of $t$, necessary 
to describe the matter behavior of the universe.\\ 
In four dimensions one obtains homogeneous space-times:
given a set of structure constants 
$C^{\alpha}_{~\beta \gamma}$ over a basis 
$\left\{ e_{\alpha} \right\}$ of the four dimensional 
Lie algebra, the ten constants 
\beq
\label{e20}
g_{\alpha \beta} = e_{\alpha} e_{\beta} \, \qquad \alpha,\beta=1,\ldots,4
\eeq
determine univocally the signature of the Lorentz 
metric. \\
As a consequence, Einstein's equations reduce 
to a system of algebraic equations for 
$g_{\alpha \beta}$ and $C^{\alpha}_{~\beta \gamma}$, 
which could in principle not be solvable for each 
transformations'
group, in view of the fact that for general formulations
with matter fields there are more constants to take 
account of. \\
In any case one needs to consider only one representative
group for each class of equivalence of the Lie groups. 
Referring to three dimensions (only the spatial 
sector), the Bianchi classification \cite{BIA97} 
determines definitely all symmetries for 
tridimensional homogeneous spaces, 
analogously to the curvature $(k=-1,0,+1)$ 
which distinguishes homogeneous and isotropic 
spaces (FRW).

\subsection{Einstein's Equations in the Synchronous Gauge}

Let's us specify the general scheme outlined above
in the specific approach of Luigi Bianchi in 1897
\cite{BIA97} and independently applied to cosmology 
by Belinski, Khalatnikov and Lifshitz in 1969
\cite{BKL70}. \\
In the cosmological approach we consider a synchronous 
reference system, so that there is an unique time 
$t$ synchronized all over the space. This can be obtained
requiring the metric tensor $g_{a b}$ to have the peculiar
diagonal form such as
\beq
\label{ay}
g_{00}=1 \, \quad g_{0 \alpha}=0 \, , \qquad \alpha=1,\ldots ,3
\eeq
in order to write the line element 
\beq
\label{bm}
{ds}^2 =N^2(t){dt}^2 -dl^2\, ,
\eeq
where $N(t)$ is the lapse function (in this section
we adopt the synchronous gauge $N\equiv 1$).  
Writing\footnote{All over this section Latin indexes
run over $0,\ldots,3$ while Greek indexes $1,\ldots,3$.} 
\beq
\label{b1}
dl^2= \gamma_{\alpha \beta}\left(t, x^{\gamma}\right)dx^{\alpha}dx^{\beta} \, ,
\eeq
$t$ is the synchronous time and the tridimensional 
tensor $\gamma_{\alpha \beta}$ describes the 
metric spatial sector. 
The identity $g_{0 \alpha}\equiv 0$ ensures the 
complete equivalence in the three spatial directions, 
while the unitarity of $g_{00}$ is always valid, 
provided an appropriate rescaling of the time coordinate.
In this reference system the time lines $(x^{\gamma}={\rm const.})$
are the geodesics of the four dimensional space-time
\beq
\frac{du^i}{ds} + \Gamma^i_{kl}u^k u^l 
=\Gamma^i_{00}=0 \, , \quad \left(i,k,l=0,\ldots, 3\right)
\label{c}
\eeq
being $u^i=\frac{dx^i}{ds}$ the tangent four vector 
to the world line $x^{\gamma}={\rm const.}$ with components 
$u^0 =1$, $u^{\alpha} =0$ and orthogonal to the 
hypersurfaces $t={\rm const.}$\footnote{This geometrical 
construction is always available but not univocally 
defined: a coordinate transformation of the type
\bea
\left\{ 
\begin{array}{ll}
t^{\prime}&=t  \\
{x^{\alpha}}^{\prime}&={x^{\alpha}}^{\prime}\left(x^{\beta}\right)
\end{array}
\right.
\label{d}
\eea
doesn't affect the time coordinate and defines the 
passage from a synchronous reference system to another one, 
fully equivalent.}. \\
Under this choice and the metric (\ref{bm}), 
the mixed components of the Einstein's equations
write 
\bea
\label{ea}
R^0_0 &=&-\frac{1}{2} \frac{\partial}{\partial t}\kappa^{\alpha}_{\alpha}  -\frac{1}{4} \kappa^{\beta}_{\alpha} \kappa^{\alpha}_{\beta} = 8\pi G\left(T^0_0 -\frac{1}{2}T\right) \\
\label{eb}
R^0_{\alpha} &=& \frac{1}{2}\left(\kappa^{\beta}_{\alpha ; \beta} -\kappa^{\beta}_{\beta ; \alpha}\right) = 8 \pi G T^0_{\alpha} \\
\label{ec}
R^{\beta}_{\alpha}&=& -P^{\beta}_{\alpha} -\frac{1}{2 \sqrt{\gamma}}\frac{\partial}{\partial t}  \left(\sqrt{\gamma}  \kappa^{\beta}_{\alpha} \right) = 8 \pi G  \left(T^{\beta}_{\alpha} -\frac{1}{2} \delta^{\alpha}_{\beta}T\right)  
\eea
being
\beq
\kappa_{\alpha \beta} =\frac{\partial \gamma_{\alpha \beta}}{\partial t} , 
\,\qquad \gamma \equiv \mid \gamma_{\alpha \beta} \mid  \, ,
\label{f}
\eeq
$T_{\alpha \beta}$ 
the momentum energy tensor for the system and 
$P_{\alpha \beta}$ the tridimensional Ricci tensor 
obtained from the metric 
$\gamma_{\alpha \beta}$\footnote{The indexes 
of these two tensors are raised and lowered 
using such reduced metric.}.\\
In (\ref{ec}) the spatial and temporal 
derivatives are split, while from (\ref{ea}) 
one obtains the property for the metric determinant
to be zero for a certain time 
(Landau-Raichoudhury theorem)\footnote{This fictitious 
singularity disappears with 
a change of reference system and is related to the 
intersection of the geodesics belonging to any arbitrary 
set with the envelope surfaces, say to the procedure
used to build a synchronous reference system.},
nevertheless the singularity with respect to the 
temporal parameter is a physical one.

\subsection{Tetradic Representation of the Metric Tensor}\label{rappresentazione}

The choice of a tetradic basis of 
four linearly independent vector fields 
(depending from the system symmetries)
permits to project all quantities in a very 
useful way to obtain new simpler equations 
to be satisfied. \\
Consider on each point of the space-time the basis 
of four linearly independent contravariant vectors 
$e^i_{\left(a\right)}$, $(a=1,\ldots,4)$, 
where the bracketed index is the tetradic one,
while $i$ is tensorial, and the covariant 
set $e_{\left(a\right) ~i} =g_{ik}e^k_{\left(a\right)}$,
being $g_{ik}$ the metric tensor. Define also the 
reciprocal set 
$e^{\left(a\right)}_i e^i_{\left(b\right)}=\delta ^a_b$
(orthogonality condition).
It is easy to check the validity of
\beq
\label{a4}
e^{\left(a\right)}_i e^k_{\left(a\right)} =\delta ^k_i \, .
\eeq
The definition is complete with
\beq
\label{a5}
e^i_{\left(a\right)} e_{\left(b\right)~i} =\eta _{ab} \, ,
\eeq
where $\eta_{ab}$ is a symmetric matrix with signature 
$(+---)$ (with the orthonormality of (\ref{a4})); 
the corresponding inverse matrix is $\eta^{ab}$,
being $\eta^{ac} \eta_{cb}=\delta^a_b$. 
Then follow the properties 
\bea
\label{a6}
\eta_{\left(a\right)\left(b\right)} e^{\left(a\right)}_i &=& e_{\left(b\right)i} \\
\eta^{\left(a\right)\left(b\right)} e_{\left(a\right)i} &=& e^{\left(b\right)}_i \\
e_{\left(a\right)i}e^{\left(a\right)}_j &=& g_{ij} ~\, .
\eea
For a generic vector $\left(A_j\right)$ or tensorial 
$\left(T_{ij}\right)$ field the tetradic components are,
in general
\bea
\label{a7}
A_{\left(a\right)}&=&e_{\left(a\right)j}A^j = e^j_{\left(a\right)}A_j \, ,\\
A^{\left(a\right)}&=&\eta^{\left(a\right)\left(b\right)}A_{\left(b\right)} =  e^{\left(a\right)}_j A^j  = e^{\left(a\right)j}A_j \, ,\\
A^i &=& e^i_{\left(a\right)}A^{\left(a\right)} = e^{\left(a\right)i}A_{\left(a\right)}
\eea
and, respectively,
\bea
\label{a8}
T_{\left(a\right)\left(b\right)}&=&e^i_{\left(a\right)}e^j_{\left(b\right)}T_{ij} = e^i_{\left(a\right)}T_{i\left(b\right)} \, ,\\
T_{ij} &=& e^{\left(a\right)}_ie^{\left(b\right)}_j T_{\left(a\right)\left(b\right)} = e^{\left(a\right)}_i T_{\left(a\right)j} \, .
\eea
The tetradic indexes can be raised or lowered with the use 
of the tensors
$\eta_{\left(a\right)\left(b\right)}$ and 
$\eta^{\left(a\right)\left(b\right)}$, while the contraction
gives then a result independent of the indexes nature.\\
By (\ref{a5}) and (\ref{a6}) one obtains also
\beq
\label{a9}
g_{ik}= e_{\left(a\right)i}e^{\left(a\right)}_k =\eta_{ab} e^{\left(a\right)}_ie^{\left(b\right)}_k
\eeq	
so that the line element becomes
\beq
\label{a10}
ds^2=\eta_{ab} \left(e^{\left(a\right)}_i dx^i\right)\left(e^{\left(b\right)}_k dx^k\right) \, .
\eeq
The choice of $\eta_{ab}$ permits to split the 
tetradic basis in one temporal and three spatial vectors. 
Nevertheless the expressions 
$dx^{\left(a\right)}=e^{\left(a\right)}_idx^i$ are not, 
in general, exact differentials of functions of the coordinates.

\subsection{Directional Derivative}\label{derivata}

The contravariant set $e_{\left(a\right)}$
of tangent vectors leads to the natural covariant derivative 
definition
\beq
\label{a11}
e_{\left(a\right)}=e^i_{\left(a\right)} \frac{\partial}{\partial x^i}
\eeq
so that the derivative of a generic scalar field $\phi$ 
along the direction $a$ is
\beq
\label{a12}
\phi_{,\left(a\right)}=e^i_{\left(a\right)} \frac{\partial \phi}{\partial x^i} = e^i_{\left(a\right)} \phi_{,i}\, .
\eeq
The general extension of such definition is
\bea
\label{a13}
A_{\left(a\right),\left(b\right)}&=&e^i_{\left(b\right)}\frac{\partial }{\partial x^i}A_{\left(a\right)}=e^i_{\left(b\right)}\frac{\partial}{\partial x^i} e^j_{\left(a\right)}A_j= \nonumber \\ 
 &=&e^i_{\left(b\right)} \nabla_{\partial _i}\left[e^j_{\left(a\right)}A_j\right] = e^i_{\left(b\right)}\left[e^j_{\left(a\right)} {A_j}_{;i} +A_k {e^k_{\left(a\right)}}_{;i}\right]
\eea
in order to rewrite
\beq
\label{a14}
A_{\left(a\right),\left(b\right)}=e^j_{\left(a\right)} A_{j;i} e^i_{\left(b\right)} +e_{\left(a\right)k;i} e^i_{\left(b\right)} e^k_{\left(c\right)} A^{\left(c\right)} \, .
\eeq
Let's introduce the Ricci's rotation 
coefficients $\gamma_{abc}$
\beq
\label{a15}
\gamma_{abc} =e_{\left(a\right)i;k}e^i_{\left(b\right)}e^k_{\left(c\right)}
\eeq
and their linear combinations
\bea
\label{a16}
\lambda_{abc} &=& \gamma_{abc} - \gamma_{acb} 
= \left(e_{\left(a\right)i;k} - e_{\left(a\right)k;i}\right) e^i_{\left(b\right)} e^k_{\left(c\right)} =\nonumber \\
&=&\left(e_{\left(a\right)i,k} -e_{\left(a\right)k,i}\right) e^i_{\left(b\right)} e^k_{\left(c\right)}
\eea
in which is has been used the identity
\beq
\label{a17}
{A_i}_{;k} -{A_k}_{;i} =\frac{\partial A_i}{\partial x^k} 
- \frac{\partial A_k}{\partial x^i} \, .
\eeq
Expression (\ref{a16}), in which the regular 
derivatives are substituted  by the covariant ones,
is invertible 
\beq
\label{a18}
\gamma_{abc} =\frac{1}{2} \left(\lambda_{abc} +\lambda_{bca}-\lambda_{cab}\right) \, .
\eeq
From the identity
\beq
\label{a19}
0=\eta_{\left(a\right)\left(b\right),i} = \left[e_{\left(a\right)k}e^k_{\left(b\right)}\right]_{;i}
\eeq
one gets the symmetry properties
\beq
\label{a20}
\gamma_{abc} =-\gamma_{bac}  \, , \qquad \lambda_{abc} =-\lambda_{acb} \, .
\eeq
Now the formalism is ready to 
find the values of the structure 
constants which leave the metric invariant under
the homogeneity constraint.\\
The basis $e_{\left(a\right)}$ permits to express 
the Lie parentheses as
\beq
\label{a21}
\left[e_{\left(a\right)}, e_{\left(b\right)}\right] =C^{\left(c\right)}_{~~\left(a\right)\left(b\right)}e_{\left(c\right)}
\eeq
where the coefficients 
$C^{\left(c\right)}_{~\left(a\right)\left(b\right)}$ 
are the 24 (in four dimensional space) 
structure constants for the group of transformations,
antisymmetric with respect to the lower indexes;
it is easy to obtain the explicit relation
\beq
\label{a22}
C^{\left(c\right)}_{~~\left(a\right)\left(b\right)} =\gamma^{\left(c\right)}_{~~\left(b\right)\left(a\right)} - \gamma^{\left(c\right)}_{~~\left(a\right)\left(b\right)} \, .
\eeq

\subsection{Ricci and Bianchi Identities}\label{identita}

By the Riemann tensor $R^m_{~ikl}$, 
the Bianchi identity becomes
\beq
\label{a23}
A_{i;k;l}-A_{i;l;k}=A_m R^m_{~ikl}
\eeq
for a generic four vector $A_i$ and, 
applied to the tetradic basis,
\beq
\label{a24}
e_{\left(a\right)i;k;l}-e_{\left(a\right)i;l;k}=e^m_{\left(a\right)} R_{mikl} \, .
\eeq
Projecting this expression on the basis itself one obtains
\bea
\label{a25}
R_{\left(a\right)\left(b\right)\left(c\right)\left(d\right)} &=&R_{mikl} e^m_{\left(a\right)}e^i_{\left(b\right)}e^k_{\left(c\right)}e^l_{\left(d\right)} = \nonumber \\
&=& -\gamma_{\left(a\right)\left(b\right)\left(c\right),\left(d\right)} + \gamma_{\left(a\right)\left(b\right)\left(d\right),\left(c\right)} + \nonumber \\
&\, &+
\gamma_{\left(a\right)\left(b\right)\left(f\right)} \left[\gamma_{\left(c\right)} {\kern 0pt}  ^{\left(f\right)} {\kern 0pt}  _{\left(d\right)} -  \gamma_{\left(d\right)} {\kern 0pt}  ^{\left(f\right)} {\kern 0pt}  _{\left(c\right)}\right] + \nonumber \\
&\, &+\gamma_{\left(f\right)\left(a\right)\left(c\right)}~\gamma_{\left(b\right)} {\kern 0pt}  ^{\left(f\right)} {\kern 0pt}  _{\left(d\right)} +  \nonumber \\
&\, &-
\gamma_{\left(f\right)\left(a\right)\left(d\right)}~\gamma_{\left(b\right)} {\kern 0pt}  ^{\left(f\right)} {\kern 0pt}  _{\left(c\right)} \, .
\eea
In view of the re-expression of the 
interesting quantities in terms of the Ricci's 
rotation coefficients 
$\gamma_{\left(a\right)\left(b\right)\left(c\right)}$ 
and subsequently in terms of the structure constants,
it is easy to see the great simplification in the 
formulas, provided a space for which exists 
a set of such constants 
$C^{\left(a\right)}_{~\left(b\right)\left(c\right)}$, 
like the analysis made by Bianchi for the homogeneous 
spaces leads to \cite{BIA97}.

The length element before the transformation is
\beq
\label{f1}
dl^2 =\gamma_{\alpha \beta} \left( x^1, x^2, x^3\right)dx^{\alpha}dx^{\beta}
\eeq 
that, under a general change of coordinates, transforms to 
\beq
\label{f2m}
dl^2 =\gamma_{\alpha \beta} \left( {x^{\prime1}}, {x^{\prime2}}, {x^{\prime3}}\right){dx^{\prime}}^{\alpha}{dx^{\prime}}^{\beta}
\eeq 
where the functional form $\gamma_{\alpha \beta}$ 
has to be the same under the homogeneity constraint. 
Such requirement, in a uniform non Euclidean space, 
leads to invariance of the three independent 
differential forms in (\ref{a10}) which are not exact 
differential of any function of the coordinates
and, in tetradic form, are
\beq
\label{f3}
e^{\left(a\right)}_{\alpha} dx^{\alpha} \, .
\eeq
The spatial invariant line element rewrites
\beq
\label{f4}
dl^2 =\eta_{ab} \left( e^{\left(a\right)}_{\alpha} dx^{\alpha} \right) \left(e^{\left(b\right)}_{\beta} dx^{\beta} \right)
\eeq
so that the metric tensor becomes
\beq
\label{f5}
\gamma_{\alpha \beta} 
=\eta_{ab} e^{\left( a \right)}_{\alpha}e^{\left(b\right)}_{\beta} \, ,
\eeq
maintaining for $\eta_{ab}$ the definition (\ref{a5}),
as a function of the time variable only.

The symmetry properties for the space determine
the specific choice of the basis vectors which, 
in general, are not orthogonal and consequently the 
metric $\eta_{ab}$ is not diagonal.\\
In pure specific case the relations between 
the three vectors are 
\bea
\label{f6}
e_{\left(1\right)} &=& \frac{1}{v} \left[ e^{\left(2 \right)} \wedge e^{\left(3 \right)} \right] \nonumber \\
e_{\left(2\right)} &=& \frac{1}{v} \left[ e^{\left(3 \right)} \wedge e^{\left(1 \right)} \right] \\
e_{\left(3\right)} &=& \frac{1}{v} \left[ e^{\left(1 \right)} \wedge e^{\left(2 \right)} \right] \nonumber 
\eea 
where $v$ is the product
\beq
\label{f7}
v= \mid e^{\left(a \right)}_{\alpha} \mid =\left( e^{\left(1\right)} \cdot \left[e^{\left(2 \right)} \wedge e^{\left(3 \right)} \right]\right)
\eeq
and where it is natural to interpret 
${\bf e}_{\left(a\right)}$ and 
${\bf e}^{\left(a\right)}$ as the Cartesian vectors
of components $e^{\alpha}_{\left(a\right)}$ and 
$e^{\left(a\right)}_{\alpha}$ respectively. \\
The metric tensor determinant (\ref{f5}) 
takes the value 
\beq
\label{f8}
\gamma=\eta v^2
\eeq	 
being $\eta$ the determinant of the matrix $\eta_{ab}$. \\
The space invariance as in (\ref{f2m}) 
is equivalent to the invariance of (\ref{f3}), then
\beq
\label{f9}
e^{\left(a\right)}_{\alpha} \left(x \right) dx^{\alpha} 
=e^{\left(a\right)}_{\alpha} \left(x^{\prime} \right) dx^{\prime \alpha}
\eeq
where $e^{\left(a\right)}_{\alpha}$ on both sides 
of the equation are the same functions of the old 
and new coordinates respectively. 
Making some algebra one gets
\beq
\label{f10}
\frac{\partial x^{\prime \beta}}{\partial x^{\alpha}} = 
e^{\beta}_{\left(a\right)} \left(x^{\prime} \right) e^{\left(a\right)}_{\alpha} \left(x \right) \, ,
\eeq
which is a system of differential equations 
to determine the functions 
$x^{\prime \beta}\left( x \right)$ on the given basis.  
The integrability of this system requires the 
satisfaction of Schwartz's conditions
\beq
\label{f11}
\frac{\partial^2 x^{\prime \beta}}{\partial x^{\alpha}\partial x^{\gamma}} =
 \frac{\partial^2 x^{\prime \beta}}{\partial x^{\gamma}\partial x^{\alpha}} \, ,
\eeq
and by explicit calculations leads to
\bea
\label{f12}
&\Big[&\frac{\partial e^{\beta}_{\left(a\right)} \left(x^{\prime} \right)}{\partial x^{\prime \delta}}  e^{\delta}_{\left(b\right)} \left(x^{\prime} \right)  -  
\frac{\partial e^{\beta}_{\left(b\right)} \left(x^{\prime} \right)}{\partial x^{\prime \delta}}  e^{\delta}_{\left(a\right)} \left(x^{\prime} \right) \Big] e^{\left(b\right)}_{\gamma} \left(x\right)e^{\left(a\right)}_{\alpha} \left(x\right)=  \nonumber \\
\, &=& e^{\beta}_{\left(a\right)} \left(x^{\prime} \right) \left[ \frac{\partial e^{\left(a\right)}_{\gamma} \left(x\right)}{ \partial x^{\alpha}} - \frac{\partial e^{\left(a\right)}_{\alpha} \left(x\right)}{ \partial x^{\gamma}}	\right] \, . 
\eea
Using the properties of the tetradic basis
and multiplying both sides of (\ref{f12}) by $e^{\alpha}_{\left(d\right)}\left(x\right)e^{\gamma}_{\left(c\right)}
\left(x\right)e^{\left(f\right)}_{\beta}\left(x^{\prime}\right)$ 
and some algebra the expression on the left
hand side becomes
\bea
\label{f13}
e^{\left(f\right)}_{\beta}\left(x^{\prime}\right) \left[\frac{\partial e^{\beta}_{\left(d\right)} \left(x^{\prime} \right)}{\partial x^{\prime \delta}}  e^{\delta}_{\left(c\right)} \left(x^{\prime} \right)- \frac{\partial e^{\beta}_{\left(c\right)} \left(x^{\prime} \right)}{\partial x^{\prime \delta}}  e^{\delta}_{\left(d\right)} \left(x^{\prime} \right) \right] = \nonumber \\
\, \quad =e^{\beta}_{\left(c\right)} \left(x^{\prime} \right) e^{\delta}_{\left(d\right)} \left(x^{\prime} \right)\left[ \frac{\partial e^{\left(f\right)}_{\beta} \left(x^{\prime}\right)}{ \partial x^{\prime \delta}} - \frac{\partial e^{\left(f\right)}_{\delta} \left(x^{\prime}\right)}{ \partial x^{\prime \gamma}}\right] \, .
\eea
Analogously one obtains for the right hand side 
an identical expression but different only for 
being function of $x$. Being the transformation 
$x\rightarrow x^{\prime}$ arbitrary, both 
expressions have to be equal to the same constant,
giving the group constant of 
structure\footnote{For the indexes of $C^c_{~ab}$ the 
parentheses are unimportant.}
\beq
\label{f14}
\left( \frac{\partial e^{\left(c\right)}_{\alpha}}{\partial x^{\beta}} 
- \frac{\partial e^{\left(c\right)}_{\beta}}{\partial x^{\alpha}}\right) e^{\alpha}_{\left(a\right)}e^{\beta}_{\left(b\right)} =C^c_{~ab} \, .
\eeq
Multiplying (\ref{f14}) by $e^{\gamma}_{\left(c\right)}$
gives the uniformity conditions over the space
\beq
\label{f15}
e^{\alpha}_{\left(a\right)} \frac{\partial e^{\gamma}_{\left(b\right)}}{\partial x^{\alpha}} - e^{\beta}_{\left(b\right)} \frac{\partial e^{\gamma}_{\left(a\right)}}{\partial x^{\beta}} =C^c_{~ab}e^{\gamma}_{\left(c\right)}\, .
\eeq
The expression on the left hand side corresponds
to the definition of $\lambda^c_{~ab}$ (\ref{a16}), 
then constants. \\
The antisymmetry property holds, 
see (\ref{a21}) or (\ref{a22}), with respect 
to the lower indexes
\beq
\label{f16}
C^c_{~ab}=-C^c_{~ba} \, .
\eeq
Such relations can be rewritten in a compact form 
in terms of the linear operators 
\beq
\label{f17}
X_a =e^{\alpha}_{\left(a\right)}  \frac{\partial }{\partial x^{\alpha}}
\eeq
so that (\ref{f15}) becomes 
\beq
\label{f18}
\left[ X_a ,X_b \right] \equiv  X_a X_b - X_b X_a =C^c_{~ab}X_c \, ,
\eeq
and the homogeneity is expressed as Jacobi identity
\beq	
\label{f19}
\left[ \left[ X_a,X_b \right] , X_c \right] + \left[ \left[ X_b,X_c \right] , X_a \right] +\left[ \left[ X_c,X_a \right] , X_b \right] =0  
\eeq
which, in terms of the structure constants, reads
\beq
\label{f20}
C^f_{~ab}C^d_{~cf}+C^f_{~bc}C^d_{~af}+C^f_{~ca}C^d_{~bf}=0 \, .
\eeq
By dual transformation one can get the two indexes
structure constants, more convenient for calculations
\beq
\label{f21}
C^c_{~ab}= \varepsilon_{abd}C^{dc}
\eeq
where $\varepsilon_{abc}=\varepsilon^{abc}$ 
is the Levi-Civita unitary antisymmetric tensor 
$(\varepsilon_{123}=+1)$. The commutation rules 
(\ref{f18}) expressed in the new constants acquire the 
compact form
\beq
\label{f22}
\varepsilon^{abc}X_b X_c =C^{ad}X_d \, .
\eeq
Antisymmetry is implied in the definition (\ref{f21}), 
while the Jacobi identity (\ref{f20}) becomes
\beq
\label{f23}
\varepsilon_{bcd} C^{cd}C^{ba}=0 \, 
\eeq
and (\ref{f15}) for the set (\ref{f21}) 
is equivalent to the vectorial form
\beq
\label{f24}
C^{ab} =-\frac{1}{v} e^{\left(a\right)} {\rm rot}~ e^{\left(b\right)} \, .
\eeq
Any linear transformation with constant coefficients
\beq
\label{f25}
e_{\left(a\right)} = A^b_a e^{\left(b\right)}
\eeq
shows the non univocal choice of the three 
reference vectors in the differential 
forms (\ref{f3}) and with respect to such 
transformations $\eta_{ab}$ and $C^{ab}$ 
behave like tensors. 
Condition (\ref{f23}) is the only one to 
be satisfied by the structure constants, 
considering only non equivalent combinations 
with respect to the transformation (\ref{f25}). \\
The classification of non equivalent homogeneous 
spaces reduces to the determination of all non 
equivalent combinations of the constants $C^{ab}$. \\
Imposing condition (\ref{f23}) one gets the relations
\bea
\label{f26}
\left[ X_1,X_2\right] &=&-aX_2 +n_3 X_3 \nonumber \\
\left[ X_2 ,X_3 \right] &=&n_1X_1 \\
\left[ X_3,X_1\right]&=& n_2X_2 +aX_3 \nonumber
\eea
where $a$ and $(n_1, n_2, n_3)$ are constants related 
to the structure constants. 
Upon reduction of all non equivalent sets under rescaling 
all possible uniform spaces can be summarized following 
the Bianchi classification as in the following table.

\begin{table}[htbp]\label{tab:bianchi}
\begin{center}
\begin{tabular}{||l||c|c|c|r||}
%
%
\textbf{Type} 	&\quad \textbf{a} \quad	&\quad $\bf{n_1}$\quad 	& \quad$\bf{n_2}$ \quad	& \quad$\bf{n_3}$ \quad 	 \\
\hline 
\hline
I	&0	&0	&0	&0	\\

II	&0	&1	&0	&0	\\

VII	&0	&1	&1	&0	\\

VI	&0	&1	&-1	&0	\\
\hline 
\hline

\textbf{IX} &\bf{0}	&\bf{1}	&\bf{1}	&\bf{1}	\\
\hline
\textbf{VIII}	&\bf{0}	&\bf{1}	&\bf{1}	&\bf{-1}	\\
\hline 
\hline
V	&1	&0	&0	&0	\\

IV	&1	&0	&0	&1	\\

VII	&a	&0	&1	&1	\\

$ 
\begin{array}{ll} 
{\rm III} ~(a=1) \\
{\rm VI} ~(a \not=1)
\end{array}  \Biggl\}$ &a	&0	&1	&-1	\\
\end{tabular}
\caption{Bianchi classification -- Non equivalent structure constants}
\end{center}
\end{table}
\section{Piecewise Representation of the Mixmaster}

\subsection{Field Equations}\label{campo}

In a synchronous reference system, the metric
for a homogeneous model writes
\beq
\label{g1}
ds^2 = dt^2 - \eta_{ab} \left(t\right) e^{\left(a \right)}_{\alpha} \left( x^{\gamma}\right)e^{\left(b \right)}_{\beta} \left(x^{\gamma}\right) dx^{\alpha}dx^{\beta}
\eeq 
where the reference vectors 
$e^{\left(a \right)}_{\alpha}$ are determined 
through (\ref{f15}), once specified the 
structure constants. 
The matrix $\eta_{ab}\left(t\right)$ describes 
the temporal evolution of the tridimensional geometry, 
to be derived from the Einstein's
field equations which reduce to an 
ordinary differential system, involving functions 
of the $t$ only, provided the projection 
of all spatial part of vectors and tensors over the 
tetradic basis chosen, using
\bea
\label{g2g}
R_{\left(a\right)\left(b\right)}&=&R_{\alpha\beta} e^{\alpha}_{\left(a\right)}e^{\beta}_{\left(b\right)} \nonumber \\
R_{0 \left(a\right)}&=&R_{0 \alpha}e^{\alpha}_{\left(a\right)} \nonumber \\
T_{\alpha\beta} &=&T_{\left(a\right)\left(b\right)}e^{\left(a\right)}_{\alpha}e^{\left(b\right)}_{\beta} \\
T_{\alpha 0} &=&T_{\left(a\right) 0}e^{\left(a\right)}_{\alpha} \nonumber \\
u^{\left(a\right)} &=&u^{\alpha}e^{\left(a\right)}_{\alpha}  \, .\nonumber
\eea
Homogeneity reflects over all scalar quantities
preventing the presence of any spatial gradient, 
incompatible with the problem symmetry. \\
The matrix $\eta_{ab}$ is the projection over 
the basis of the spatial metric 
$\gamma_{\alpha \beta}$ and the role of $\eta_{ab}$ 
and $\eta^{ab}$ to raise and lower the indexes is clear. \\
The projection of the field equations 
(\ref{ea})-(\ref{ec}) over the tetrad gives 
\bea
\label{g3a}
R^0_0 &=&-\frac{1}{2} {\dot{\kappa}}^{\left(a\right)}_{\left(a\right)} -\frac{1}{4}{\kappa}^{\left(b\right)}_{\left(a\right)}{\kappa}^{\left(a\right)}_{\left(b\right)} \\
\label{g3b}
R^0_{\left(a\right)}&=&-\frac{1}{2}{\kappa}^{\left(c\right)}_{\left(b\right)}\left(C^b_{~ca} -\delta^b_a C^d_{~dc}\right) \\
\label{g3c}
R^{\left(a\right)}_{\left(b\right)}&=&-\frac{1}{2 \sqrt{\eta}} \frac{\partial}{\partial t}{\left(\sqrt{\eta} {\kappa}^{\left(b\right)}_{\left(a\right)}\right)} -P^{\left(a\right)}_{\left(b\right)}
\eea
where 
\bea
\label{g4}
{\kappa}_{\left(a\right)\left(b\right)} &=&\dot{\eta}_{\left(a\right)\left(b\right)} \nonumber \\
{\kappa}^{\left(b\right)}_{\left(a\right)} &=&\dot{\eta}_{\left(a\right)\left(c\right)}\eta^{\left(c\right)\left(b\right)}
\eea
and the dot is the derivative with respect 
to $t$ and the tridimensional 
Ricci tensor projected
\bea
\label{g5}
P_{\left(a\right)\left(b\right)}=\eta_{\left(b\right)\left(c\right)}P^{\left(c\right)}_{\left(a\right)}
\eea
in terms of the structure constants becomes
\bea
\label{g6}
P_{\left(a\right)\left(b\right)}=
&-&\frac{1}{2} \Big(C^{cd}_{~~b}C_{cda}+ C^{cd}_{~~b}C_{dca}+  \nonumber \\
\, &-& \frac{1}{2} C_b^{~cd}C_{acd} +C^c_{~cd}C_{ab}^{~d} +C^c_{~cd}C_{ba}^{~~d}\Big)  \, .
\eea
Moreover it is not necessary the explicit 
form of the vector basis
as functions of the coordinate for . 
This task is related to the symmetry degree of the 
models considered.\\

\subsection{Kasner Solution}\label{kasner}

The simplest homogeneous cosmological model
corresponds to the Bianchi type I, whose 
structure constants identically vanish, 
implying also the Ricci tensor components 
to be zero so that 
\beq
\label{g7}
\begin{array}{ll} 
e^a_{\alpha} =\delta^a_{\alpha} \\
C^c_{ab} \equiv 0
\end{array}  \Biggl\} \Longrightarrow P_{ab} =0 \, .
\eeq
Under this conditions and in empty space, 
equations (\ref{g3a})-(\ref{g3c}) reduce to the system 
\bea
\label{g8a}
{\dot{\kappa}}^a_a +\frac{1}{2} {\kappa}^b_a{\kappa}^a_b &=&0  \\
\label{g8b}
\frac{1}{\sqrt{\gamma}} \frac{\partial}{\partial t}{\left( \sqrt{\gamma} {\kappa}^b_a\right)} &=&0 \, .
\eea 
By (\ref{g8b}) one gets the first integral  
\beq
\label{g9}
\sqrt{\gamma}\kappa^b_a =2 \lambda^b_a ={\rm const.}\, .
\eeq
The contraction of the $a$ and $b$ indexes gets
\beq
\label{g10}
\kappa^a_a = \frac{\dot{\gamma}}{\gamma}=\frac{2}{\sqrt{\gamma}}\lambda^a_a \, ,
\eeq
and then  
\beq
\label{g11}
\gamma= G t^2 \, ,\qquad G={\rm const.} \, .
\eeq
Without loss of generality, upon a coordinates 
rescaling, put such constant equal to one
and then also
\beq
\label{g12}
\lambda^a_a =1 \, .
\eeq
Substituting (\ref{g9}) in (\ref{g8a}) one finds 
the relation for the constants $\lambda^b_a$
\beq
\label{g13}
\lambda^a_b\lambda^b_a =1 \, .
\eeq 
Lowering index $b$ in (\ref{g9}) gives 
a system of ordinary differential equations
with respect to $\gamma_{ab}$
\beq
\label{g14}
\dot{\gamma}_{ab} =\frac{2}{t} \lambda^c_a\lambda_{cb} \, .
\eeq
The set of $\lambda^c_a$ can be considered as a 
matrix for a linear transformation 
and by an appropriate change 
of coordinates $(x^1,x^2,x^3)$ is reducible to 
a diagonal form. Given $p_1, p_2, p_3$ as the 
corresponding matrix eigenvalues, 
taken real and different, in correspondence 
of the normalized eigenvectors 
${\bf n}^{\left(1\right)},{\bf n}^{\left(2\right)},{\bf n}^{\left(3\right)}$,
the solution of (\ref{g14}) is 
\beq
\label{g15}
\gamma_{ab} =t^{2p_1}n^{\left(1\right)}_a n^{\left(1\right)}_b + t^{2p_2}n^{\left(2\right)}_a n^{\left(2\right)}_b+t^{2p_3}n^{\left(3\right)}_a n^{\left(3\right)}_b
\eeq
where the constant coefficients for the powers 
of $t$ can be reduced to unity, once rescaled the
coordinates.
If the tetrad vectors are parallel to the 
coordinate axes, say $(x,y,z)$, 
metric reduces to
\beq
\label{g16}
ds^2 =dt^2 -t^{2p_1}dx^2-t^{2p_2}dy^2-t^{2p_3}dz^2 \, .
\eeq
Constants $p_1, p_2, p_3$ are three arbitrary 
numbers, called Kasner indexes, which have to satisfy 
the conditions 
\bea
\label{g17}
p_1+p_2+p_3&=&1 \\
\label{g18}
p^2_1+p^2_2+p^2_3&=&1 \,  ,
\eea
as a direct consequence of (\ref{g14}) and (\ref{g13}).
Except the cases $(0,0,1)$ and 
$(-\frac{1}{3},\frac{2}{3},\frac{2}{3})$, 
Kasner indexes have to be different and one of 
them can acquire negative value.
Given the order 
\beq
\label{g19}
p_1 <p_2<p_3 \, ,
\eeq
the corresponding variation interval is 
\bea
\label{g20}
-\frac{1}{3} &\leq & p_1 \leq 0 \nonumber \\
0 &\leq & p_2 \leq \frac{2}{3} \\
\frac{2}{3} &\leq & p_3 \leq 1 \nonumber \, .
\eea
These numbers can also be represented in 
the parametric form 
\bea
\label{g20a}
p_1\left(u\right) &=&\frac{-u}{1+u+u^2} \nonumber \\
p_2\left(u\right) &=&\frac{1+u}{1+u+u^2} \\
p_3\left(u\right) &=&\frac{u\left(1+u\right)}{1+u+u^2} \nonumber
\eea
where the parameter $u$ varies in the 
interval $1 \leq u < + \infty$.
When $u<1$, the parameterization can be 
reduced to the same variability 
interval for $p_1, p_2, p_3$, holding  
\bea
\label{g20c}
p_1 \left(\frac{1}{u}\right) &=&p_1\left(u\right) \nonumber \\
p_2 \left(\frac{1}{u}\right) &=&p_3\left(u\right)  \\
p_3 \left(\frac{1}{u}\right) &=&p_2\left(u\right). \nonumber
\eea
As functions of $u$, the parameters 
$p_1\left(u\right)$ and $p_3\left(u\right)$ 
monotonically increase, while 
$p_2\left(u\right)$ monotonically decreases.
This metric corresponds to a flat space, 
even anisotropic, where the volume grows
proportionally with increasing $t$ 
and distances along $(y,z)$ axes increase 
while along $x$ decrease.
The time $t=0$ is the singular point for the 
solution and such singularity is not avoidable under
any reference system change, while the invariant
quantities of the curvature tensor diverge
except the case $p_1=p_2=0,~p_3=1$ in which
the metric is reducible to the Galilean form
once posed the parameterization
\bea
\label{g21}
t \sinh x^3 &=&\xi \nonumber \\
t \cosh x^3 &=&\tau \, .
\eea 
The metric (\ref{g16}) is the exact solution for the 
Einstein's equations in empty space but, close the a 
singular point, say small $t$, it represents
an approximate solution. \\
A generalized solution coincides only with an
approximation of the metric, in the sense that 
dominant terms of such metric as powers of $t$ 
have analogous form to (\ref{g16}). \\
In general, given a synchronous reference system, 
the metric can be expressed in the form (\ref{bm}), 
where the spatial line element $dl$ is
\beq
\label{g22}
dl^2 =\left( a^2l_{\alpha}l_{\beta}+b^2m_{\alpha}m_{\beta}+c^2n_{\alpha}n_{\beta} \right)dx^{\alpha}dx^{\beta}
\eeq
once posed
\bea
\label{g23}
a&=&t^{p_l} \nonumber \\
b&=&t^{p_m}  \\
c&=&t^{p_n} \nonumber 
\eea
and the three tridimensional vectors 
${\bf l},{\bf m},{\bf n}$ redefine the directions
along which the spatial lengths vary following 
the power laws (\ref{g23}). Such quantities are
functions of the spatial coordinates. 
In view of the fact that exponents (\ref{g23}) 
are all different, the spatial metric (\ref{g22}) 
is always anisotropic. \\
An eventual presence of matter doesn't affect 
the generality of these conclusions and can be 
introduced in the metric (\ref{g22})-(\ref{g23}) 
via four coordinates functions to determine the 
initial distribution of matter and the three initial 
velocity components. The behavior of matter in the 
vicinity of a singular point is determined by equations
of motion in a given gravitational field, following 
classical hydrodynamics.

\subsection{Oscillatory Regime}\label{oscillatorio}

Differently from the Kasner solution,
in all other homogeneous models the projection 
of the tridimensional Ricci tensor has components
different from zero and the complexification induced 
prevents an analytical description of the solution, 
except for the Bianchi type II model.\\
In the following is described the asymptotic behavior 
of the homogeneous models in the vicinity of a 
singular point.\\
The most general and interesting case relies in the 
Bianchi types VIII and IX models (Mixmaster) \cite{M69}, 
while others can be considered as simplifications. \\
The general solution is, by definition, stable. 
A perturbation to the system is equivalent to a change 
in the initial conditions at a given time, but the general 
solution satisfies arbitrary initial conditions, 
then the given perturbation cannot affect the form 
of the solution.

The structure constants sets considered are (see table)
\bea
\label{g25}
&{\rm Bianchi ~VIII} : \quad C^{11} =C^{22}=1, \, C^{33}=-1 \nonumber \\
&{\rm Bianchi ~IX }: \quad C^{11} =C^{22}=1=C^{33}=1 \, .
\eea
Let's consider the case of Bianchi IX. 
If the matrix $\eta_{ab}$ is diagonal, 
the components $R^0_{\left(a\right)}$ of the Ricci 
tensor in (\ref{g3a})-(\ref{g3c}) vanish identically
in the synchronous reference system and, by (\ref{g6}),
the same holds for the off diagonal components of 
$P_{\left(a\right)\left(b\right)}$. 
The remaining Einstein's equations give for 
the functions $a\left(t\right),b\left(t\right),c\left(t\right)$ 
the system 
\bea
\label{g26}
\frac{{\left(\dot{a}bc \right)}^{\bullet}}{abc}&=&\frac{1}{2a^2b^2c^2}\left[{\left(b^2-c^2\right)}^2-{a}^4 \right] \nonumber \\
\frac{{\left(a\dot{b}c \right)}^{\bullet}}{abc}&=&\frac{1}{2a^2b^2c^2}\left[{\left(a^2-c^2\right)}^2-{b}^4 \right]  \\
\frac{{\left(ab\dot{c} \right)}^{\bullet}}{abc}&=&\frac{1}{2a^2b^2c^2}\left[{\left(a^2-b^2\right)}^2-{c}^4 \right] \nonumber 
\eea
and 
\beq
\label{g27}
\frac{\ddot{a}}{a}+\frac{\ddot{b}}{b}+\frac{\ddot{c}}{c}=0
\eeq
depending on the time variable only and 
$(~)^{\bullet}=\frac{d}{dt}$. 
Equations (\ref{g26})-(\ref{g27}) are exact and valid
also far from the singularity in $t=0$. 
Introducing the quantities
\bea
\label{g28} 
a&=&e^{\alpha} \nonumber \\
b&=&e^{\beta} \\
c&=&e^{\gamma} \nonumber
\eea
and the temporal variable $\tau$
\beq
\label{g29}
dt=abc ~d\tau
\eeq
one obtains the simplified form
\bea
\label{g30}
2\alpha_{\tau\tau}&=&{\left(b^2-c^2\right)}^2-{a}^4 \nonumber \\
2\beta_{\tau \tau}&=&{\left(a^2-c^2 \right)}^2-{b}^4 \\
2\gamma_{\tau \tau}&=&{\left(a^2-b^2\right)}^2-{c}^4 \nonumber \\
\label{g30a}
\frac{1}{2}\left(\alpha+\beta+\gamma\right)_{\tau\tau}&=&\alpha_{\tau}\beta_{\tau}+\alpha_{\tau}\gamma_{\tau}+\beta_{\tau}\gamma_{\tau}
\eea
where the index $\tau$ refers to the derivative with 
respect to $\tau$. The system (\ref{g30}), by (\ref{g30a}), 
admits the first integral 
\bea
\label{g31}
\alpha_{\tau}\beta_{\tau}&+&\alpha_{\tau}\gamma_{\tau}
+\beta_{\tau}\gamma_{\tau}= \nonumber \\
\, &=&\frac{1}{4}\Big(a^4 + b^4 +c^4 -2a^2b^2-2a^2c^2-2b^2c^2\Big)
\eea
containing only first derivatives. 
The Kasner regime (\ref{g23}) is a solution 
of (\ref{g30}) if all right hand side expressions can 
be neglected. Such a situation cannot hold indefinitely
(for $t\rightarrow0$), because there is always an 
increasing term. Suppose the case that, 
for a certain time interval, some expressions 
in (\ref{g30}) are negligible (Kasnerian regime): 
for example, the negative exponent corresponds to 
the function $a\left(t\right)$, so that $p_l=p_1$, 
and the perturbation to the given regime is due to 
terms as $a^4$, meanwhile other terms decrease.
Keeping the dominant terms in (\ref{g30}), one reduces 
to the approximate
\bea
\label{g32}
\alpha_{\tau\tau}&=&-\frac{1}{2}e^{4\alpha} \nonumber \\
\beta_{\tau\tau}&=&\gamma_{\tau\tau}=\frac{1}{2}e^{4\alpha} \, .
\eea
The solution for those equations describes the evolution 
of the metric, given the initial condition (\ref{g23}). 
Without loss of generality, once put 
\bea
\label{g33}
p_l&=&p_1 \nonumber \\
p_m&=&p_2 \\
p_n&=&p_3\nonumber 
\eea
so that 
\bea
\label{g33b}
a&\sim & t^{p_1} \nonumber \\
b&\sim & t^{p_2} \\
c&\sim & t^{p_3} \nonumber
\eea
one obtains 
\bea
\label{g34}
abc&=&\Lambda t \nonumber \\
\tau&=&\frac{1}{\Lambda}\ln t+{\rm const.}
\eea
where $\Lambda$ is a constant while initial conditions
for (\ref{g32}) are
\bea
\label{g35}
\alpha_{\tau}&=&p_1 \nonumber \\
\beta_{\tau}&=&p_2 \\
\gamma_{\tau}&=&p_3 \nonumber,
\eea
and finally
\bea
\label{g36}
\alpha&=&\Lambda p_1 \nonumber \\
\beta&=&\Lambda p_2 \\
\gamma&=&\Lambda p_3 \nonumber \, .
\eea
The first of (\ref{g32}) has the same form as 
the equation of the unidimensional motion for 
a point particle in an exponential potential 
barrier, where $\alpha$ has the role of a coordinate. 
By this analogy, the initial Kasnerian regime 
corresponds to a free motion with constant velocity 
$\alpha_{\tau}=p_1$. After the bounce on the barrier
the particle follows its free motion with the same speed
but with opposite sign $\alpha_{\tau}=-p_1$. 
Using initial conditions (\ref{g36}), the system (\ref{g32}) 
integrates as
\bea
\label{g37}
a^2 &=&\frac{2\mid p_1\mid\Lambda}{\cosh\left(2\mid p_1\mid\Lambda\tau\right)} \nonumber \\
b^2&=&{b_0}^2\exp\left[ 2\Lambda \left(p_2-\mid p_1\mid\right)\tau\right]\cosh\left(2\mid p_1\mid\Lambda \tau\right)\\
c^2&=&{c_0}^2\exp\left[ 2\Lambda \left(p_3-\mid p_1\mid\right)\tau\right]\cosh\left(2\mid p_1\mid\Lambda \tau\right)\nonumber
\eea
where $b_0$ and $c_0$ are integration constants. 
For the solutions (\ref{g37}), in the limit 
$\tau\rightarrow\infty$, the asymptotic form 
coincides with (\ref{g33b}). For 
$\tau\rightarrow -\infty$, say toward the singularity,
the asymptotic forms become
\bea
\label{g38}
a&\sim & \exp\left[\Lambda p_1 \tau\right] \nonumber \\
b&\sim & \exp \left[ \Lambda \left( p_2 +2p_1\right)\tau\right] \nonumber \\
c&\sim & \exp \left[\Lambda \left(p_3+2p_1\right)\tau\right] \\
t&\sim & \exp\left[ \Lambda \left(1+2p_1\right)\tau\right] \nonumber
\eea
and writing $a,b,c$ as functions of $t$ one gets again 
a Kasner behavior, a new Kasner {\it epoch}, 
\bea
\label{g39}
a&\sim & t^{p^{\prime}_l} \nonumber \\
b&\sim & t^{p^{\prime}_m}  \\
c&\sim & t^{p^{\prime}_n} \nonumber \\
abc&=&\Lambda^{\prime}t \nonumber
\eea
being
\bea
\label{g40}
p^{\prime}_l&=&~~\frac{\mid p_1\mid}{1-2\mid p_1\mid} \nonumber \\
p^{\prime}_m&=&-\frac{2\mid p_1\mid -p_2}{1-2\mid p_1\mid} \\
p^{\prime}_n&=&~~\frac{p_3-2\mid p_1\mid}{1-2\mid p_1\mid} \nonumber \, ,
\eea
together with
\beq
\label{g41}
\Lambda^{\prime} =\left(1- 2\mid p_1\mid \right)\Lambda \, .
\eeq
The perturbation causes the transition from a Kasner 
epoch to another in such a way that the negative power 
of $t$ passes from the direction $\textbf{l}$ 
to $\textbf{m}$, i.e. if initially $p_l$ is negative,  
in the new solution one gets ${p_m}^{\prime}<0$. 
The initial perturbation given by terms such 
$e^{4\alpha}$ is reduced substantially to zero. 
For the following evolution it will become the dominant
term in another right hand side term and follow 
a completely analogous analysis. \\
By the parameterization (\ref{g20a}) and the property 
(\ref{g20c}), the passages from a Kasner epoch to another 
re-express in a suggestive way such that if 
\bea
\label{g42}
p_l&=&p_1\left(u\right) \nonumber \\
p_m&=&p_2\left(u\right)  \\
p_n&=&p_3\left(u\right) \nonumber
\eea
then
\bea
\label{g43}
p^{\prime}_l&=&p_2\left(u-1\right) \nonumber \\
p^{\prime}_m&=&p_1\left(u-1\right)  \\
p^{\prime}_n&=&p_3\left(u-1\right) \nonumber
\eea
and is called {\it BKL map}. \\
During this transition, the function 
$a\left(t\right)$ gets a maximum value, while 
$b\left(t\right)$ a minima. After that $b$ starts
increasing, $a$ decreases and $c\left(t\right)$
holds its monotonic decrease. \\
During the change of Kasner epoch, the bigger of the 
two positive exponents maintains its character 
following (\ref{g43}). The following changes of epochs 
continue until the integer part of the initial value 
of $u$ becomes smaller then one.
After that, following (\ref{g20c}) the value 
of $u<1$ corresponds to another $u^{\prime}>1$. \\
This reparameterization realizes a new exchange in 
the two positive indexes among the corresponding 
directions, say one of the oscillating functions 
becomes decreasing and vice versa. \\
The time interval in which one scale factor 
holds monotonically decreasing while the other two
oscillate is sadi {\it Kasner era}. The sequence 
of the axis exchange and the order in which two er\ae 
of different time duration follows assumes a 
stochastic character.

\subsection{BKL Map}\label{mappa}

Each Kasner era holds for an included number of epochs 
equivalent to the integer part of the corresponding 
initial value of the parameter $u$. Being arbitrary 
and in general irrational, the regimes' alternation 
continues indefinitely, while for an exact solution 
the meaning of the exponents $p_1, p_2, p_3$ loses 
the role as in the Kasner er\ae. Such internal indetermination 
takes out meaning to the study of privileged sets of values.\\
The hypothesis underlying this asymptotic regime is based on the 
possibility to find a time interval during which it is 
possible to neglect the projection of the tridimensional 
Ricci tensor. \\
The presence of matter as a perfect fluid, described by an 
opportune momentum energy tensor, doesn't affect the 
characteristics of the regime toward the singularity.
Nevertheless the description would get a peculiar evolution 
for the matter. 
In such a case, by the hydrodynamic 
equations, the temporal evolution of the energy density 
for each Kasner epoch becomes
\beq
\label{g44}
\epsilon \sim t^{-2\left(1-p_3\right)}
\eeq 
where $p_3$ is the bigger among the set $p_1,p_2,p_3$. 
$\epsilon$ monotonically increases with the temporal variable 
decrement and diverges for $t=0$, as confirmation of 
the intrinsic singularity. \\
In all other Bianchi models the asymptotic regime 
toward the singularity is overall simplified: 
the oscillatory behavior disappears because the  
right hand side members of (\ref{g30}) vanish and 
the corresponding Kasner epoch is the only one 
describing the system evolution toward the singularity.

Oscillation amplitudes for $\alpha$ 
and $\beta$ are growing during the evolution 
of an assigned era, independently of the era chosen,
provided obviously the oscillating behavior.\\
In parallel to that, also the Kasner epochs' duration is 
increasing and the matter density increases 
monotonically as 
\beq
\label{g48}
\frac{\epsilon^{\prime}_0}{\epsilon_0}\sim {A_0}^{2k}
\eeq
being $A_0$ the original oscillation amplitude 
and $k$ the index for the $k$-th era and it 
is accelerating in the following one
\beq
\label{g49}
\frac{\epsilon^{\prime \prime}_0}{\epsilon^{\prime}_0}\sim {A_0}^{\prime 2k^{\prime}}\sim{A_0}^{2k^2 k^{\prime}}
\eeq
and so on, showing how rapid is the density 
matter growth. \\
The ongoing series of oscillations accumulate in 
the vicinity of the singular point.
Between any final instant of the universal time $t$
and $t=0$ there is an infinite number of oscillations. 
The temporal evolution is given by the natural variable
$\ln t$ and not the synchronous time $t$, going to 
$-\infty$ toward the singularity. 

The study of the iterative properties of the BKL map 
(\ref{g43}) requires another fundamental property of 
the parameter $u$.
The $s$-th corresponds to a succession of 
decreasing values of the parameter $u$ of the 
form $u^{\left(s\right)}_{max}$ (staring era), 
$u^{\left(s\right)}_{max}-1,u^{\left(s\right)}_{max}-2,\ldots,u^{\left(s\right)}_{min}$.  
One can distinguish 
\beq
\label{g45a}
u^{\left(s\right)}=k^{\left(s\right)}+x^{\left(s\right)}
\eeq
with notation 
\bea
\label{g45b}
u^{\left(s\right)}_{min}&=&x^{\left(s\right)}<1 \nonumber \\
u^{\left(s\right)}_{max}&=&k^{\left(s\right)}+x^{\left(s\right)} \, ,
\eea
$u^{\left(s\right)}_{max}$ is the maximum value of $u$, $k^{\left(s\right)}=\left[u^{\left(s\right)}_{max}\right]$
and\footnote{The square brackets 
denote the integer part function.} 
$x^{\left(s\right)}$ is the fractional part. 
The succession refers to a decreasing sequence of 
values of $u$. \\
The first Kasner era contains $k^{\left(s\right)}$ 
epochs while the subsequent one, parameterized by 
\bea
\label{g46}
u^{\left(s+1\right)}_{max}&=& \frac{1}{x^{\left(s\right)}} \nonumber \\
k^{\left(s+1\right)}&=&\left[\frac{1}{x^{\left(s+1\right)}}\right] \, ,
\eea
corresponds to the sequence of $k^{\left(s+1\right)}$. 
If the entire arrangement starts from 
$k^{\left(0\right)}+x^{\left(0\right)}$, 
the lengths $k^{\left(1\right)}, k^{\left(2\right)}, \ldots$ 
are the numbers involved in the expansion of 
$x^{\left(0\right)}$ in the continuous fraction 
\beq
\label{g47}
x^{\left(0\right)}=\displaystyle\frac{1}{k^{\left(1\right)}
+\displaystyle\frac{1}{k^{\left(2\right)}+\frac{1}{k^{\left(3\right)}+\ldots}	}} \, ,
\eeq
which is finite if corresponds to the expansion 
of a rational number but in general infinite when
the initial value is irrational. \\
In any infinite sequence $u$ constructed by (\ref{g46}) 
there are always arbitrarily small numbers 
$x^{\left(s\right)}$, but different from zero, to which 
correspond arbitrarily long er\ae $~k^{\left(s+1\right)}$.
The value assumed by all terms of such expansion is 
finite and limited and the set of all $x^{\left(0\right)}<1$ 
with this property has null measure in the interval $(0,1)$. \\
Depending on the strong dependence of the BKL map of 
the initial conditions the infinite sequence of $\left(k^{\left(0\right)},k^{\left(1\right)},k^{\left(2\right)}, \ldots \right)$ 
assumes a chaotic behavior which requires an appropriate 
analysis. 

\subsection{Statistical Description}\label{statistica}

Kasner er\ae ~iteration, the specific stochasticity 
in the parameters sequence and the iterative dynamics
require a statistical description. \\
If the initial conditions don't affect the 
model evolution, expression (\ref{g47}) is 
very sensitive to the initial values of $u$ 
and following the system along many er\ae ~it is 
possible to determine a specific probability 
distribution.\\
Any small change on the initial value of 
$u^{\left(0\right)}$ implies a sequence of numbers
$k$ completely different once explicited all the 
fraction terms, then, approaching the singularity,
one gets a stationary probability distribution 
for the values assumed by the integer part $k$ 
and by the fractional one $x$ referring to $u$. \\
The random nature of the process under which 
the sequence $k^{\left(s\right)}$ acquires asymptotically 
a stochastic character raises from the transition mechanism 
between different er\ae ~(\ref{g46}). \\
Instead of an initial value as in 
(\ref{g45a}) with $s=0$, let's consider a distribution 
for $x^{\left(0\right)}$ over the range $(0,1)$, 
described by the probability distribution 
$W_0\left(x\right)$ for $x^{\left(0\right)}=x$ to 
be in such interval. 
Given ~$w_s\left(x\right)dx$ ~the probability for 
the last term of the $s$-th series $x^{\left(s\right)}=x$ 
to be in the interval $dx$, the corresponding last 
term in the set has to be in between $\frac{1}{k+1}$ and 
$\frac{1}{k}$, necessary for the duration of 
the $s$-th series to be $k$. 
In this case, the probability for such duration is 
given by 
\beq
\label{g50}
W_s\left(k\right)=\int^{\frac{1}{k}}_{\frac{1}{1+k}} w_{s-1}\left(x\right)dx \, ,
\eeq
and the recurrence formula relating the probability 
distribution $w_{s+1}\left(x\right)$ with 
$w_{s}\left(x\right)$ is
\beq
\label{g51}
w_{s+1}\left(x\right) =\sum^{\infty}_{k=1}\frac{1}{{\left(k+x\right)}^2}w_s\left(\frac{1}{k+x}\right) \, .
\eeq
The recursive definition (\ref{g51}) generates $w_{s+n}$ 
($n$ generic integer) which, for increasing $s$, tends
to a stationary probability distribution $w\left(x\right)$ 
where the initial conditions are disappeared
\beq
\label{g52}
w\left(x\right) =\sum^{\infty}_{k=1}\frac{1}{{\left(k+x\right)}^2}w\left(\frac{1}{k+x}\right) \, ,
\eeq
whose normalized solution of (\ref{g52}) is 
\beq
\label{g53}
w\left(x\right)=\frac{1}{\left(1+x\right)\ln 2} \, .
\eeq
Once given the probability distribution for $x$, 
the corresponding series length $k$ is derivable as
\beq
\label{g54}
W\left(k\right)=\int^{\frac{1}{k}}_{\frac{1}{1+k}} w\left(x\right)dx=\frac{1}{\ln 2}\ln{\frac{{\left(k+1\right)}^2}{k\left(k+2\right)}} \, .
\eeq
Finally, being $k$ and $x$ not independent, 
they have to admit a stationary probability distribution 
correlated 
\beq
\label{g55}
w\left(k,x\right)=\frac{1}{\left(k+x\right)\left(k+x+1\right)\ln2}
\eeq
which, given $u=k+x$, rewrites as a stationary distribution 
for the parameter $u$ 
\beq
\label{g56}
w\left(u\right) =\frac{1}{u\left(u+1\right)\ln2} \, .
\eeq
This is the basic result for the study of 
the model evolution statistical properties, 
for the homogeneous cosmological model considered. \\
The perturbative term of the Kasner regime
in the Einstein's equations is identified as the negative 
power law and the corresponding scale factor.
Nevertheless, for $u\gg 1$ in any epoch,
the parameterization (\ref{g20a}) provides Kasner 
indexes in the 
asymptotic form 
\bea
\label{g57}
p_1&\cong &\frac{1}{u} \nonumber \\
p_2&\cong &\frac{1}{u} \\
p_3 &\cong & 1- \frac{1}{u^2} \nonumber
\eea
where $p_1,p_2$ have absolute values comparable and 
both close to zero. In this scheme, the perturbation 
is produced equally by terms such $t^{4p_1}$ and 
$t^{4p_2}$. Such situation provides equations whose 
solution, far from the singularity, has an oscillatory 
behavior. \\
The obtained relations for the statistical dynamics
during a Kasner era loose their validity when the system
acquires a small oscillations regime, due to its strong 
instability.

\section{Covariant Approach to the Mixmaster Chaos}

\subsection{The Hamiltonian Formulation}

In order to provide a Hamiltonian formulation 
of the Mixmaster dynamics, we rewrite the line element 
as follows \cite{MTW}
\begin{equation} 
\label{a} 
ds^2=-{N(\eta)}^2d{\eta}^2+e^{2\alpha}\left(e^{2\beta}\right)_{ij}\sigma^i \sigma^j 
\end{equation} 
where $N(\eta)$ denotes the lapse function, $\sigma^i$ are the dual 1-forms associated with the anholonomic basis \footnote{ The dual 1-forms of the considered models are given by: \\[1em]

(Bianchi ~VIII): $\left\{ 
\begin{array}{lll} 
\sigma ^1 =-\sinh \psi \sinh\theta d\phi~ + ~\cosh \psi d\theta \\
\sigma ^2= -\cosh \psi \sinh\theta d\phi ~+~\sinh \psi d\theta \\
\sigma ^3=~\cosh\theta d\phi ~+ ~d \psi 
\end{array} \right.$ 
\\[1em]

(Bianchi ~IX): \quad  $\left\{ 
\begin{array}{lll} 
\sigma ^1 = ~\sin \psi \sin\theta d\phi ~+~\cos \psi d\theta \\
\sigma ^2 = -\cos \psi \sin\theta d\phi ~+~\sin \psi d\theta \\
\sigma ^3 = ~\cos\theta d\phi ~+~d \psi  
\end{array} \right.$
}
and $\beta_{ij}$ is a traceless $3\times 3$ symmetric matrix  ${\rm diag}(\beta_{11},\beta_{22},\beta_{33})$; $\alpha$, $N$, $\beta_{ij}$ are functions of $\eta$ only. Parameterizing the matrix $\beta_{ij}$ by the usual Misner variables \cite{M69}
\begin{eqnarray}
\label{a1}
\beta_{11}&=&\beta_+ + \sqrt3 \beta_- \nonumber \\
\beta_{22}&=&\beta_+ - \sqrt3 \beta_-   \\
\beta_{33}&=&-2 \beta_+ \nonumber 
\end{eqnarray} 
the dynamics of the Mixmaster model is described by a canonical variational principle

\begin{equation} 
\label{b} 
\delta I=\delta\int L\, d\eta=0 \, ,
\end{equation} 
with Lagrangian $L$ 

\begin{equation} \label{r} 
L=\frac{6 D}{N}\left[{-{\alpha}^{\prime}}^2+{{\beta_+}^{\prime}}^2+{{\beta_-}^{\prime}}^2\right]- \frac{N}{D}V\left(\alpha, \beta_+ , \beta_-\right) \, .
\end{equation}  
Here ${()}^{\prime} = \frac{d}{d\eta}$, $D\equiv \det e^{\alpha +\beta_{ij}}=e^{3\alpha}$  and the potential $V\left(\alpha, \beta_+ , \beta_-\right)$ reads
\bea
\label{a2v} 
V=\frac{1}{2} \Big(&~& D^{4H_1}+D^{4H_2}+D^{4H_3}\Big) +\nonumber \\
&~& -D^{2H_1 +2H_2}\pm D^{2H_2 +2H_3}\pm D^{2H_3 +2H_1} \, ,
\eea  
where $(+)$ and $(-)$ refers respectively to Bianchi type VIII and IX, and the anisotropy parameters $H_i ~(i=1,2,3)$ denote the functions \cite{KM97}
\begin{eqnarray}
\label{ssaa}
H_1 &=& \frac{1}{3}+ \frac{\beta_+ + \sqrt{3} \beta_-}{3 \alpha} \nonumber \\
H_2 &=& \frac{1}{3}+ \frac{\beta_+ - \sqrt{3} \beta_-}{3 \alpha}   \\
H_3 &=&\frac{1}{3}- \frac{2\beta_+}{3 \alpha}  \nonumber \, .
\end{eqnarray} 
In the limit $D\rightarrow 0$ the second three terms of the above potential turn out to be negligible with respect to the first one.
Let's introduce the new (Misner-Chitr\'e-like) variables 
\begin{eqnarray}
\label{f2}
&\alpha =& -e^{f\left(\tau\right)}\xi \nonumber \\
&\beta_+ =& e^{f\left(\tau\right)}\sqrt{\xi^2 -1}\cos \theta \\
&\beta_- =& e^{f\left(\tau\right)}\sqrt{\xi^2 -1}\sin \theta  \, ,\nonumber 
\end{eqnarray} 
with $f$ denoting a generic functional form of $\tau$, $1\le \xi <\infty$ and $0\le \theta < 2 \pi$. Then the Lagrangian (\ref{r}) reads
\bea 
\label{g2}
L=\frac{6 D}{N} \Big[ \frac{{\left(e^f {\xi}^{\prime}\right)}^2}{\xi ^2 -1} 
&+&{\left(e^f {\theta}^{\prime}\right)}^2\left(\xi ^2 -1\right) 
-{{\left(e^f\right)}^{\prime}}^2 \Big]+ \nonumber \\
\,  &-&\frac{N}{D}V \left( f\left(\tau\right), \xi, \theta \right) \, .
\eea 
In terms of $f\left(\tau\right)$, $\xi$ and $\theta$ we have
\begin{equation} 
D= exp\left\{ -3 \xi \cdot e^{f\left(\tau\right)} \right\}
\label{h} 
\end{equation} 
and since $D \rightarrow 0$ toward the singularity, independently of its particular form, in this limit $f$ must approach infinity.
The Lagrangian (\ref{r}) leads to the conjugate momenta 
\begin{eqnarray}
\label{i}
p_{\tau}&=&\frac{\partial L}{\partial {\tau}^{\prime}} = -\frac{12 D}{N}{\left(e^f \cdot \frac{df}{d\tau}\right)}^2  {\tau}^{\prime}   \nonumber \\
p_{\xi}&=&\frac{\partial L}{\partial {\xi}^{\prime}} = \frac{12 D}{N} \frac{e^{2f}}{{\xi}^2 -1}{\xi}^{\prime}  \\
p_{\theta}&=&\frac{\partial L}{\partial {\theta}^{\prime}} = \frac{12 D}{N}e^{2f}\left({\xi}^2 -1\right)  {\theta}^{\prime} \nonumber 
\end{eqnarray} 
which by a Legendre transformation make the variational principle (\ref{b}) assume the Hamiltonian form  
\begin{equation} 
\delta \int \left(   p_{\xi} {\xi}^{\prime} +  p_{\theta} {\theta}^{\prime}+    p_{\tau}  {\tau}^{\prime} - \frac{Ne^{-2f}}{24 D} {\cal H}                            \right) d\eta =0 \, ,
\label{m} 
\end{equation} 
being 
\begin{equation} 
{\cal H} = -\frac{{p_{\tau}}^2}{\left(\frac{df}{d\tau}\right)^2} + 
 {p_{\xi}}^2\left(\xi ^2 -1\right) +\frac{{p_{\theta}}^2}{\xi ^2 -1} +24 V e^{2f}  \, .
\label{n} 
\end{equation}

\subsection{Reduced Variational Principle}

By variating (\ref{m}) with respect to $N$ we get the constraint ${\cal H} =0$, which solved provides 
\begin{equation} 
-p_{\tau}\equiv \frac{df}{d\tau}\cdot {\cal H}_{ADM} = \frac{df}{d\tau} \cdot \sqrt{\varepsilon ^2 +24 V e^{2f}}
\label{n2} 
\end{equation}
where
\begin{equation}
\varepsilon ^2 = \left({\xi}^2 -1\right){p_{\xi}}^2 +\frac{{p_{\theta}}^2}{{\xi}^2 -1} 
\label{d2} 
\end{equation} 
in terms of which the variational principle (\ref{m}) reduces to 
\begin{equation} 
\delta \int \left(   p_{\xi} {\xi}^{\prime} +  p_{\theta} {\theta}^{\prime} - {f}^{\prime}{\cal H}_{ADM} \right) d\eta =0 \, .
\label{q} 
\end{equation} 
Since the equation for the temporal gauge actually reads
\begin{equation} 
N\left(\eta\right)= \frac{12 D}{{\cal H}_{ADM}} e^{2f}  \frac{df}{d\tau}  {\tau}^{\prime} \, ,
\label{rs} 
\end{equation} 
our analysis remains fully independent of the choice of the time variable until the form of $f$ and ${\tau}^{\prime}$ is not fixed.

The variational principle (\ref{q}) provides  the Hamiltonian equations for ${\xi}^{\prime}$ and ${\theta}^{\prime}$ 
\footnote{In this study the corresponding equations for $p^{\prime}_{\xi}$ and $p^{\prime}_{\theta}$ are not relevant.}
\begin{eqnarray}
{\xi}^{\prime}&=& \frac{f^{\prime}}{{\cal H}_{ADM}}\left(\xi ^2 -1\right)p_{\xi}           \nonumber \\
{\theta}^{\prime}&=& \frac{f^{\prime}}{{\cal H}_{ADM}} \frac{p_{\theta}}{\left(\xi ^2 -1\right)} \, .
\label{s}
\end{eqnarray}
Furthermore can be straightforward derived the important relation
\bea 
\frac{d\left({\cal H}_{ADM}f^{\prime}\right)}{d\eta} &=& \frac{\partial \left({\cal H}_{ADM}f^{\prime}\right)}{\partial\eta} \Longrightarrow \nonumber \\
\, &~&\Longrightarrow \frac{d\left({\cal H}_{ADM}f^{\prime}\right)}{df} = \frac{\partial \left({\cal H}_{ADM}f^{\prime}\right)}{\partial f} \, ,
\label{t} 
\eea 
i.e. explicitly
\begin{equation} 
\frac{\partial{\cal H}_{ADM}}{\partial f}=
\frac{e^{2f}}{2 {\cal H}_{ADM}} 24\cdot \left( 2V+ \frac{\partial V}{\partial f} \right) \, .
\label{u} 
\end{equation} 
In this reduced Hamiltonian formulation, the functional $f\left(\eta\right)$ plays simply the role of a parametric function of time and actually the anisotropy parameters $H_i$ $(i=1,2,3)$ are functions of the variables $\xi, \theta$ only 

\begin{eqnarray} 
\label{v4}
H_1 &=& \frac{1}{3} - \frac{\sqrt{\xi ^2 - 1}}{3\xi }\left(\cos\theta + \sqrt{3}\sin\theta \right)  \nonumber \\ 
H_2 &=& \frac{1}{3} - \frac{\sqrt{\xi ^2 - 1}}{3\xi }\left(\cos\theta - \sqrt{3}\sin\theta \right)  \\ 
H_3 &=& \frac{1}{3} + 2\frac{\sqrt{\xi ^2 - 1}}{3\xi } \cos\theta \, .  \nonumber 
\end{eqnarray}

Finally, toward the singularity ($D \rightarrow 0$ i.e. $f \rightarrow \infty$) by the expressions (\ref{a2v}, \ref{h}, \ref{v4}), we see that \footnote{By $O()$ we mean terms of the same order of the inclosed ones.}
\begin{equation}
\frac{\partial V}{\partial f} = O\left(e^f V\right) \, .
\label{z0}
\end{equation} 
Since in the domain $\Gamma_H$, where all the $H_i$ are simultaneously greater than 0, the potential term $U\equiv e^{2f} V$ can be modeled by the potential walls

\begin{eqnarray}
\label{aa}
U_\infty = 
&\Theta _\infty \left(H_1\left(\xi, \theta\right)\right) + \Theta _\infty \left(H_2\left(\xi, \theta\right)\right) + \Theta _\infty \left(H_3\left(\xi, \theta\right)\right) \, \\ 
 &\Theta _\infty \left(x\right) = \left\{ 
\begin{array}{lll} 
+ \infty & {\rm if} & x < 0 \\ 
\quad 0 & {\rm if} & x >  0 
\end{array} \right.\nonumber
\end{eqnarray}
therefore in $\Gamma_H$ the ADM Hamiltonian becomes (asymptotically) an integral of motion
\[
\forall \{\xi, \theta\}\in{ \Gamma_H}  \nonumber
\]
\beq
\label{bb}
\left\{ 
\begin{array}{lll} 
{\cal H}_{ADM}= \sqrt{\varepsilon ^2 +24\cdot U} \cong \varepsilon =E ={\rm const.} \\ 
\displaystyle  \frac{\partial {\cal H}_{ADM}}{\partial f} 
=\frac{\partial E}{\partial f} =  0  \, .
\end{array}
\right.
\eeq

The key point for the use of the Misner-Chitr\'e-like variables relies on the independence of the time variable for the anisotropy parameters $H_i$. 

\subsection{The Jacobi Metric and the Billiard Representation}

Since above we have shown that asymptotically to the singularity ($f \rightarrow \infty$, i.e. $\alpha \rightarrow  -\infty$) $d{\cal H}_{ADM}/df =0$ i.e. ${\cal H}_{ADM} =\epsilon =E={\rm const.}$, the variational principle (\ref{q}) reduces to
\bea
\label{cc}
\delta \int \Big( p_{\xi} d\xi + p_{\theta} d\theta &-&Edf \Big) = \nonumber \\
&=&\delta \int \left(  p_{\xi} d\xi + p_{\theta} d\theta \right)=0 \, ,
\eea
where we dropped the third term in the left hand side since it behaves as an exact differential.

By following the standard Jacobi procedure \cite{A89} to reduce our variational principle to a geodesic one, we set ${x^a}^{\prime} \equiv g^{ab}p_b$, and by the Hamiltonian equation (\ref{s}) we obtain the metric
\begin{eqnarray}
g^{\xi \xi} &=&\frac{f^{\prime}}{E}\left({\xi}^2 -1\right) \nonumber \\
g^{\theta \theta} &=&\frac{f^{\prime}}{E} \frac{1}{{\xi}^2 -1} \, .
\label{dd}
\end{eqnarray}
By these and by the fundamental constraint relation
 \begin{equation}
\left({\xi}^2 -1\right){p_{\xi}}^2 +\frac{{p_{\theta}}^2}{{\xi}^2 -1} =E^2 \, ,
\label{ee}
\end{equation}
we get 
\begin{equation}
g_{ab}{x^a}^{\prime} {x^b}^{\prime} =\frac{f^{\prime}}{E} \left\{ \left({\xi}^2 -1\right){p_{\xi}}^2 +\frac{{p_{\theta}}^2}{{\xi}^2 -1}\right\}=f^{\prime}  E \, .
\label{ff}
\end{equation}
By the definition ${x^a}^{ \prime}= \frac{dx^a}{ds} \frac{ds}{d\eta}\equiv u^a \frac{ds}{d\eta}$, (\ref{ff}) rewrites
\begin{equation}
g_{ab}u^a u^b \left( \frac{ds}{d\eta} \right) ^2  = f^{\prime} E \, ,
\label{gg}
\end{equation}
which leads to the key relation
\begin{equation}
d\eta = \sqrt{ \frac{g_{ab}u^a u^b}{f^{\prime} E }}~ds \, .
\label{hh}
\end{equation}
Indeed expression (\ref{hh}) together with $p_{\xi} \xi^{\prime} +p_{\theta} \theta^{\prime}=Ef^{\prime}$ allows us to put the variational principle (\ref{cc}) in the geodesic form
\bea
\delta \int  f^{\prime} E ~d\eta  
&=& \delta \int  \sqrt{ g_{ab}u^a u^b f^{\prime} E} ~ds= \nonumber \\
&~&\qquad \quad= \delta \int  \sqrt{ G_{ab}u^a u^b}~ds =0
\label{ii}
\eea
where the metric $G_{ab} \equiv f^{\prime} E g_{ab}$ satisfies the normalization condition $G_{ab}u^a u^b =1$ and therefore \footnote{We take the positive root since we require that the curvilinear coordinate $s$ increases monotonically with increasing value of $f$, i.e. approaching the initial cosmological singularity.} 
\begin{equation}
\frac{ds}{d\eta}=Ef^{\prime}\Rightarrow \frac{ds}{df} =E \, .
\label{ll}
\end{equation}
Summarizing, in the region $\Gamma_H$ the considered dynamical problem reduces to a geodesic flow on a two dimensional Riemannian manifold described by the line element
\begin{equation}
ds^2 =E^2 \left[ 	\frac{d{\xi }^2}{{\xi}^2 -1}+  
\left(\xi^2 -1\right) d {\theta }^2 \right] \, .
\label{mm}
\end{equation}
Now it is easy to check that the above metric has negative curvature, since the associated curvature scalar reads $R=-\frac{2}{E^2}$;
therefore the point-universe moves over a negatively curved bidimensional space on which the potential wall (\ref{a2v}) cuts the region $\Gamma_{H}$. By a way completely independent of the temporal gauge we provided a satisfactory representation of the system as isomorphic to a billiard on a Lobachevsky plane \cite{A89}.

\subsection{Invariant Lyapunov Exponent}

In order to characterize the dynamical instability of the billiard in terms of an invariant treatment (with respect to the choice of the coordinates $\xi$, $\theta$), let us introduce the following (orthonormal) tetradic basis
\begin{eqnarray} 
v^i &=&\left(\frac{\sqrt{{\xi}^2-1}}{E}, 0\right)         \nonumber \\
w^i &=&\left(0,\frac{1}{E \sqrt{{\xi}^2-1}}\right)       \,  .
\label{nn}
\end{eqnarray} 
Indeed the vector $v^i$ is nothing else than the geodesic field, i.e. 
\begin{equation}
\frac{Dv^i}{ds}=\frac{dv^i}{ds}+\Gamma^i_{kl}v^k v^l =0 \, ,
\label{oo}
\end{equation}
while the vector $w^i$ is parallel transported along the geodesic, according to the equation
\begin{equation}
\frac{Dw^i}{ds}=\frac{dw^i}{ds}+\Gamma^i_{kl}v^k w^l =0 \, ,
\label{pp}
\end{equation}
where by $\Gamma^i_{kl}$ we denoted the Christoffel symbols constructed by the metric (\ref{mm}).
Projecting the geodesic deviation equation along the vector $w^i$ (its component along the geodesic field $v^i$ does not provide any physical information about the system instability), the corresponding connecting vector (tetradic) component $Z$ satisfies the following equivalent equation 
\begin{equation}
\frac{d^2 Z}{ds^2}=\frac{Z}{E^2} \, .
\label{qqq}
\end{equation}
This expression, as a projection on the tetradic basis, is a scalar one and therefore completely independent of the choice of the variables.
Its general solution reads
\begin{equation}
Z\left(s\right)=c_1 e^{\frac{s}{E}}+c_2 e^{-\frac{s}{E}} \, , \qquad c_{1,2}={\rm const.} \,  ,
\label{rr}
\end{equation}
and the invariant Lyapunov exponent defined as \cite{PE77} 
\begin{equation}
\lambda_v =\sup \lim_{s\rightarrow \infty} \frac{\ln\left(Z^2+ \left(\frac{dZ}{ds}\right)^2\right)}{2s} \, ,
\label{ss}
\end{equation}
in terms of the form (\ref{rr}) takes the value
\begin{equation}
\lambda_v =\frac{1}{E} > 0 \, .
\label{tt}
\end{equation}
When the point-universe bounces against the potential walls, it is reflected from a geodesic to another one thus making each of them unstable. Though up to the limit of our potential wall approximation, this result shows without any ambiguity that, independently of the choice of the temporal gauge, the Mixmaster dynamics is isomorphic to a well described chaotic system. Equivalently, in terms of the BKL representation, the free geodesic motion corresponds to the evolution during a Kasner epoch and the bounces against the potential walls to the transition between two of them. By itself, the positive Lyapunov number (\ref{tt}) is not enough to ensure the system chaos, since its derivation remains valid for any Bianchi type model; the crucial point is that for the Mixmaster (type VIII and IX) the potential walls reduce the configuration space to a compact region ($\Gamma_H$), ensuring that the positivity of $\lambda_v$ implies a real chaotic behavior (i.e. the geodesic motion fills the entire configuration space).

Summarizing, our analysis shows that for any choice of the time variable, we are able to give the above stochastic representation of the Mixmaster model, provided the factorized coordinate transformation in the configuration space
\begin{eqnarray}
\label{uu}
&\alpha &= -e^{f\left(\tau\right)} a\left(\theta , \xi\right) \nonumber\\
&\beta_+ &=~e^{f\left(\tau\right)} b_+ \left(\theta , \xi\right) \\
&\beta_- &=~e^{f\left(\tau\right)} b_- \left(\theta , \xi\right) \, , \nonumber 
\end{eqnarray}
where $f,a,b_{\pm}$ denote generic functional forms of the variables $\tau, \theta, \xi$.

It is worth noting that the success of our analysis, in showing the time gauge independence of the Mixmaster chaos, relies on the use of a standard ADM reduction of the variational principle (which reduces the system by one degree of freedom) and overall because, adopting Misner-Chitr\'e-like variables, the asymptotic potential walls are fixed in time. The difference between our approach and the one presented in \cite{SL90} (see also for a critical analysis \cite{BT93}) consists effectively in these features, though in those works is even faced the problem of the Mixmaster chaos covariance with respect to the choice of generic configuration variables.

\section{Statistical Mechanics Approach}

\subsection{Covariance of the Mixmaster Invariant Measure}

In order to reformulate the description of the Mixmaster stochasticity
in terms of the Statistical Mechanics \cite{IM02}, we adopt in (\ref{q}) the restricted time gauge $\tau^{\prime}=1$, leading to the 
variational principle 
\begin{equation} 
\delta \int \left(   p_{\xi} \frac{d\xi }{df } +  p_{\theta} 
\frac{d\theta}{df } 
- {\cal H}_{ADM} \right) df = 0 .
\label{px} 
\end{equation} 
In spite of this restriction, for any assigned time variable $\tau$ (i.e. $\eta$) there  
exists a corresponding function $f \left(\tau \right)$ 
(i.e. a set of Misner-Chitr\'e-like variables) defined by the (invertible) relation 
\begin{equation} 
\frac{df }{d\tau} = \frac{{\cal H}_{ADM}}{12 D }N\left(\tau \right) e^{-2f } . 
\label{qx} 
\end{equation} 
As a consequence of the variational principle (\ref{px}) we have again the expression (\ref{u}).

In agreement with this scheme, in the region $\Gamma_H$ where the potential vanishes,
we have by (\ref{qx}) $d{\cal H}_{ADM}/df = 0$, i.e. 
$\varepsilon = E = {\rm const.}$ (by other words the ADM Hamiltonian 
approaches an integral of motion). 

Hence the analysis to derive the invariant measure for the system follows 
the same lines presented in \cite{KM97,M00}. 

Indeed we got again a representation of the Mixmaster dynamics in terms of 
a two-dimensional point-universe moving within closed potential walls and over a 
negative curved surface (the Lobachevsky plane \cite{KM97}), described by the line element (\ref{mm}).
Due to the bounces against 
the potential walls and to the instability of the geodesic flow on such a plane, the dyna\-mics acquires a stochastic 
feature. This system, 
admitting an ``energy-like'' constant of motion $\left(\varepsilon = E\right)$,  
is well-described by a {\it microcanonical ensemble}, whose 
Liouville invariant measure reads 
\begin{equation} 
d\varrho = A \delta \left(E - \varepsilon \right)d\xi d\theta dp_{\xi }dp_{\theta }  \, , \qquad A={\rm const.}
\label{ux} 
\end{equation} 
where $\delta\left(x\right)$ denotes the Dirac function.
After the natural positions 
\begin{equation} 
p_{\xi } = \frac{\varepsilon}{\sqrt{\xi ^2 - 1}}\cos\phi \, , \qquad p_{\theta } = \varepsilon \sqrt{\xi ^2 - 1}\sin\phi  \, , 
\label{v} 
\end{equation} 
being $0\le\phi\le 2\pi$, 
and the integration over all 
possible values of $\varepsilon$ (depending on the initial 
conditions, they do not contain any 
information about the system chaos), 
we arrive to the uniform invariant measure 
\begin{equation} 
\label{xp}
d\mu = d\xi d\theta d\phi \frac{1}{8\pi ^2} \, . 
\end{equation} 

The validity of our potential approximation is legitimated by implementing 
the Landau-Raichoudhury theorem 
\footnote{Such a theorem, 
within the mathematical assumptions founding Einsteinian dynamics, states that 
in a synchronous reference it always exists a given instant of time in 
correspondence to which the metric determinant vanishes monotonically.} 
near the initial singularity (placed by convention in $T = 0$, 
where $T$ denotes the synchronous time, i.e. $dT=- N\left(\tau\right)d\tau$), we easily get 
that $D$ vanishes monotonically (i.e. for $T\rightarrow 0$ we 
have $d \ln D/dT > 0$). 
In terms of the adopted time variable $\tau$ 
$\left(D\rightarrow 0 \Rightarrow f(\tau) \rightarrow \infty\right)$, we have
\begin{equation} 
\frac{d \ln D}{d \tau } = 
\frac{d \ln D}{d T}\frac{dT}{d\tau } = 
- \frac{d \ln D}{d T}N\left(\tau \right)  
\label{x1} 
\end{equation}
and therefore $D$ vanishes monotonically for increasing $\tau$ as 
soon as, by (\ref{qx}), $d\Gamma / d\tau >0$. 

Now the key point of our analysis is that any stationary solution of the Liouville theorem, like (\ref{u}), remains valid 
for any choice of the time variable $\tau$; indeed in \cite{M00} the construction of the Liouville theorem with respect to the 
variables $(\xi, \theta,\phi)$ shows the existence of such properties even for the invariant measure (\ref{x}).

We conclude remarking how, when approaching the singularity $f\rightarrow~\infty$ (i.e. ${\cal H}_{ADM}\rightarrow E$), the time gauge relation (\ref{qx}) simplifies to
\begin{equation} 
\label{qq}
\frac{df }{d\tau} = 
\frac{E e^{-2f +3\xi e^{f}}}{12 }N\left(\tau \right) e^{-2f} ; 
\label{x2} 
\end{equation} 
in agreement with the analysis presented in \cite{M00}, during a free geodesic motion the asymptotic functions 
$\xi\left(f\right), \theta\left(f\right), \phi\left(f\right)$, are provided 
by the simple system
\bea
\frac{d\xi}{df}&=&\sqrt{\xi^2-1}\cos\phi \nonumber \\ 
\frac{d\theta}{df}&=&\frac{\sin\phi}{\sqrt{\xi^2-1}} \nonumber \\ 
\frac{d\phi}{df}&=&-\frac{\xi\sin\phi}{\sqrt{\xi^2-1}} \, . 
\eea
Once getting $\xi\left(f\right)$ as the parametric solution
\begin{eqnarray}
\label{xx}
\xi \left(\phi\right)&=&\frac{\rho}{\sin^2\phi} \nonumber \\
f \left( \phi \right)&=& 
\frac{1}{2} {\rm arctanh} \left(\frac{1}{2} \frac{\sin^2\phi 
+a^2 \left(1+\cos^2\phi \right)}{a\rho \cos \phi}\right) 
 +b \nonumber \\
\rho&\equiv&\sqrt{a^2 +\sin^2\phi} \, \quad a,b={\rm const.}\in \Re 
\end{eqnarray}
it reduces, for a free geodesic motion, equation (\ref{qq}) to a simple
differential one for the function $f\left(\tau\right)$. 

However, the global behavior of $\xi$ along the whole geodesic flow, is described
by the invariant measure (\ref{x}) and therefore relation (\ref{qq}) takes a stochastic character. 
If we assign one of the two functions $f \left(\tau\right)$ or $N\left(f\right)$ 
analytically, the other one acquires a stochastic behavior.
We see how the one-to-one correspondence between any lapse function $N\left(\eta\right)$ and the associated 
set of Misner-Chitr\`e-like variables, ensures the covariant nature with respect to the time-gauge of the Mixmaster universe stochastic behavior.\\

\subsection{Quantum Nature of the Mixmaster Chaos}

In what follows we will consider the particular case
$f(\tau)\equiv \tau$ and to make clear its asymptotic nature,
we redefine the invariant measure as follows
\begin{equation} 
d\mu = w_{\infty}\left(\xi, \theta,\phi\right) d\xi d\theta d\phi 
\equiv \frac{1}{8\pi ^2}d\xi d\theta d\phi   \, .
\label{x} 
\end{equation} 
Summarizing, over the reduced phase space\footnote{$S^1_{\phi}$ denotes the $\phi$-circle.} 
$\{\xi,\theta\}\otimes S^1_{\phi}$ the distribution $w_{\infty}$ 
behaves like the step-function 
\begin{equation}
\label{step}
w_{\infty}\left(\xi, \theta, \phi\right)=\left\{ \begin{array}{lll} \displaystyle
\frac{1}{8\pi^2} \quad &\forall& \left\{ \xi, \theta, \phi\right\} \in \Gamma_H\otimes S^1_{\phi}  \\ 
0  \quad &\forall& \left\{ \xi, \theta, \phi\right\} \not\in \Gamma_H\otimes S^1_{\phi} 
\end{array} 
\right. \, .
\end{equation}
Once performed the transformation (\ref{v}) over the 
Hamiltonian equations (\ref{s}), in the asymptotic limit for which
$U\rightarrow U_{\infty}\Rightarrow \varepsilon=E={\rm const.}$, we get in 
$\Gamma_H$ the free geodesic motion \cite{M00}
\bea
\label{ham}
\frac{d\xi}{d\tau}&=&\sqrt{\xi^2-1}\cos\phi \nonumber \\ 
\frac{d\theta}{d\tau}&=&\frac{\sin\phi}{\sqrt{\xi^2-1}} \nonumber \\ 
\frac{d\phi}{d\tau}&=&-\frac{\xi\sin\phi}{\sqrt{\xi^2-1}} \, . 
\eea

This dynamical system induces the {\it stationary} continuity 
equation for the distribution function $w_{\infty}(\xi, \theta, \phi)$ 
describing the ensemble representation
\bea
\label{cont}
\sqrt{\xi^2-1}\cos\phi \frac{\partial w_{\infty}}{\partial \xi}&+&
\frac{\sin\phi}{\sqrt{\xi^2-1}}\frac{\partial w_{\infty}}{\partial \theta}+ \nonumber \\
&-& \frac{\xi\sin\phi}{\sqrt{\xi^2-1}}\frac{\partial w_{\infty}}{\partial \phi}=0 \, .
\eea
If now we restrict our attention to the distribution function on the 
configuration space $\Gamma_H$
\begin{equation}
\label{conf}
\varrho\left(\xi, \theta\right)\equiv \int_0^{2\pi}w_{\infty}\left(\xi, \theta, \phi\right)d\phi \, ,
\end{equation}
by (\ref{cont}) we get for such restricted form the two dimensional 
continuity equation 
\begin{equation}
\label{cont2}
\sqrt{\xi^2-1}\cos\phi \frac{\partial \varrho_{\infty}}{\partial \xi}+
\frac{\sin\phi}{\sqrt{\xi^2-1}}\frac{\partial \varrho_{\infty}}{\partial \theta}=0 \, .
\end{equation}
The {\it microcanonical} solution on the whole configuration 
space $\{\xi, \theta\}$ reads 
\begin{equation}
\label{steprho}
\varrho_{\infty}\left(\xi, \theta\right)=\left\{ \begin{array}{lll} \displaystyle
\frac{1}{4\pi} \qquad &\forall& \left\{ \xi, \theta\right\} \in \Gamma_H  \\ 
0  \qquad &\forall& \left\{ \xi, \theta\right\} \not\in \Gamma_H 
\end{array} 
\right. \, .
\end{equation}

The main result of \cite{IM01} and \cite{IM01a,IM01b}, 
is the proof that the chaos of the
Bianchi IX model above outlined is an intrinsic feature of its dynamics and 
not an effect induced by a particular class of references: in fact the whole
MCl formalism and its consequences can be developed in a framework 
free from the choice of a specific time gauge. \\
Since this intrinsic chaos appears close enough to the Big Bang,
we infer that it has some relations with the indeterministic 
quantum dynamics the model performs in the {\it Planckian era}. 
This relation between quantum and deterministic chaos is searched
in the sense of a semiclassical limit for a canonical 
quantization of the model \cite{IM02}.

The asymptotical principle (\ref{q}) describes a two dimensional 
anholonomic Hamiltonian system, which can 
be quantized by a natural Schr\"oedinger approach 
\begin{equation}
\label{sch}
i \hbar \frac{\partial \psi}{\partial \tau}=\hat{{\cal H}}_{ADM}\psi \, ,
\end{equation}
being $\psi=\psi(\tau,\xi,\theta)$ the wave function for the point-universe
and, implementing $\hat{{\cal H}}_{ADM}$ (see (\ref{bb})) to 
an operator\footnote{The only non vanishing canonical commutation relations are
\[ \left[ \hat{\xi},\hat{p_{\xi}}\right] =i\hbar \, , \qquad \left[ \hat{\theta},\hat{p_{\theta}}\right] =i\hbar \, .
\]}, i.e.
\begin{eqnarray}
\label{op}
\xi &\rightarrow &\hat{\xi} \, , \qquad \qquad  \quad \qquad  \theta \rightarrow \hat{\theta} \, ,  \nonumber \\
p_{\xi} &\rightarrow & \hat{p_{\xi}} \equiv -i \hbar \frac{\partial}{\partial \xi} \, , \qquad 
p_{\theta} \rightarrow \hat{p_{\theta}} \equiv -i \hbar \frac{\partial}{\partial \theta} \, ,
\end{eqnarray} 
the equation (\ref{sch}) rewrites explicitly, in the asymptotic limit $U\rightarrow U_{\infty}$,
\begin{eqnarray}
\label{sch1}
i \frac{\partial \psi}{\partial \tau} &=& \sqrt{\hat{\varepsilon}^2 +\frac{U_{\infty}}{\hbar^2}}~\psi 
= \Big[-\sqrt{\xi^2 -1}\frac{\partial }{\partial \xi} \sqrt{\xi^2 -1} \frac{\partial }{\partial \xi} + \nonumber \\
&-& \frac{1}{\sqrt{\xi^2 -1}}\frac{\partial }{\partial {\theta}}\frac{1}{\sqrt{\xi^2 -1}}\frac{\partial }{\partial {\theta}} +\frac{U_{\infty}}{\hbar^2}\Big]^{1/2} \psi \, ,
\end{eqnarray}
where we took an appropriate symmetric normal ordering prescription 
and we left $U_{\infty}$ to stress that the potential cannot be neglected 
on the entire configuration space $\{\xi, \theta\}$ and, being infinity out 
of $\Gamma_H$, it requires as boundary condition for $\psi$
to vanish outside the potential walls
\begin{equation}
\label{bound}
\psi\left(\partial \Gamma_H\right)=0 \, .
\end{equation}
The {\it quantum} equation (\ref{sch1}) is equivalent to the 
Wheeler-DeWitt one for the same Bianchi model, once separated the positive and negative
frequencies solutions \cite{KU81}, with the advantage that now $\tau$ 
is a real time variable\footnote{This equivalence can be easily checked by taking 
the square of the operators on both sides of the equation.}.
Since the potential walls $U_{\infty}$ are time independent, 
a solution of this equation has the form
\begin{equation}
\label{sol}
\psi\left(\tau, \xi, \theta\right)= \sum_{n=1}^{\infty} c_n e^{-i E_n \tau / \hbar} \varphi_n\left(\xi, \theta\right)
\end{equation}
where $c_n$ are constant coefficients and we assumed a 
discrete ``energy'' spectrum because the quantum point-universe is restricted in the 
finite region $\Gamma_H$ and the position (\ref{sol}) in (\ref{sch1}) leads 
to the eigenvalue problem 
\begin{eqnarray}
\label{eig}
\Big[-\sqrt{\xi^2 -1}&~&\frac{\partial }{\partial \xi} \sqrt{\xi^2 -1} \frac{\partial }{\partial \xi} +\nonumber \\
&-&\frac{1}{\sqrt{\xi^2 -1}}\frac{\partial }{\partial {\theta}}\frac{1}{\sqrt{\xi^2 -1}}
\frac{\partial }{\partial {\theta}} \Big] \varphi_n = \nonumber \\
&~&= {\left(\frac{{E_n}^2-U_{\infty}}{\hbar^2}\right)} \varphi_n 
\equiv \frac{{E_{\infty}}^2_n}{\hbar^2} \varphi_n     \, . 
\end{eqnarray} 
In what follows we search the semiclassical solution of this equation 
regarding the eigenvalue ${E_{\infty}}_n$ as a finite constant 
(i.e. we consider the potential walls as finite) and only 
at the end of the procedure we will take the limit for $U_{\infty}$ (\ref{aa}).

We infer that in the semiclassical limit when $\hbar \rightarrow 0$ and the 
{\it occupation number} $n$ tends to infinity (but $n\hbar$ approaches a finite value)
the wave function $\varphi_n$ approaches a function $\varphi$ as 
\begin{equation}
\label{phi}
\mathop {\lim }\limits_{\scriptstyle n \to \infty  \hfill \atop 
\scriptstyle \hbar  \to 0 \hfill}
\varphi_n\left(\xi, \theta\right) =\varphi\left(\xi, \theta\right) \, , \qquad
\mathop {\lim }\limits_{\scriptstyle n \to \infty  \hfill \atop 
\scriptstyle \hbar  \to 0 \hfill}
{E_{\infty}}_n ={E_{\infty}} \, .
\end{equation}
The expression $\varphi$ is taken as a semiclassical expansion up to 
the first order, i.e. 
\begin{equation}
\label{expa}
\varphi\left(\xi, \theta \right) =\sqrt{r \left(\xi, \theta \right)} 
\exp\left\{ i\frac{S\left(\xi, \theta \right)}{\hbar}\right\} \, ,
\end{equation}
where $r$ and $S$ are functions to be determined. \\
Substituting (\ref{expa}) in (\ref{eig}) and separating the 
real from the complex part we get two independent equations, i.e.
\begin{eqnarray}
\label{1eq}
\displaystyle
{E_{\infty}}^2&=& \underbrace{\left( \xi^2-1\right) \left( \frac{\partial S}{\partial \xi}\right)^2 + 
\frac{1}{\xi^2-1} \left( \frac{\partial S}{\partial \theta}\right)^2 }_{\rm classical~term} +  \nonumber \\
&-& \frac{\hbar^2}{\sqrt{r}} \left[ \sqrt{\xi^2-1}\frac{\partial}{\partial \xi}\sqrt{\xi^2-1}\frac{\partial}{\partial \xi} + 
\frac{1}{\xi^2-1}\frac{\partial^2}{\partial \theta ^2} \right]\sqrt{r}    
\end{eqnarray}
where we multiplied both sides by $\hbar^2$ and, respectively,
\begin{equation}
\label{2eq}
\underbrace{\sqrt{\xi^2-1} \frac{\partial }{\partial \xi} \left( \sqrt{\xi^2-1}~r \frac{\partial S}{\partial \xi} \right) 
+\frac{1}{\xi^2-1}\frac{\partial }{\partial \theta} \left( r \frac{\partial S}{\partial \theta}\right)}_{O(1/\hbar)} =0 \, .
\end{equation}
In the limit $\hbar \rightarrow 0$ the second term of (\ref{1eq}) is negligible
meanwhile the first one reduces to the Hamilton-Jacobi equation 
\begin{equation}
\label{hamb}
\left( \xi^2-1\right) \left( \frac{\partial S}{\partial \xi}\right)^2 + 
\frac{1}{\xi^2-1} \left( \frac{\partial S}{\partial \theta}\right)^2={E_{\infty}}^2 \, .
\end{equation} 
The solution of (\ref{hamb}) can be easily checked to be 
\beq
\label{esse}
S\left(\xi, \theta\right) = \int\left\{ \frac{1}{\sqrt{\xi^2 -1}}\sqrt{E^2_{\infty}
 - \frac{k^2}{\xi^2-1}}~d\xi +k ~d\theta\right\} \, ,
\eeq
where $k$ is an integration constant.\\
We observe that (\ref{hamb}), through the identifications 
\begin{equation}
\label{ide}
\frac{\partial S}{\partial \xi} =p_{\xi} \, , \, \frac{\partial S}{\partial \theta}=p_{\theta} 
\quad\Longleftrightarrow \quad S=\int\left( p_{\xi} d\xi + p_{\theta} d \theta \right)    \, ,
\end{equation} 
is reduced to the algebraic relation
\begin{equation}
\label{hamr}
\left( \xi^2-1\right) {p_{\xi}}^2 + \frac{1}{\xi^2-1} {p_{\theta}}^2={E_{\infty}}^2 \, .
\end{equation}
The constraint (\ref{hamr}) is nothing more than the asymptotic one 
${\cal H}_{ADM}^2=E^2={\rm const.}$ and can be solved setting 
\bea
\label{sethm}
\frac{\partial S}{\partial \xi} &=& p_{\xi} \equiv \frac{E_{\infty}}{\sqrt{\xi^2-1}}\cos \phi \, , \nonumber \\
\frac{\partial S}{\partial \theta}&=&p_{\theta}\equiv E_{\infty}\sqrt{\xi^2-1}\sin \phi \, ,
\eea
where $\phi\in [0,2\pi[$ is a momentum-function related 
to $\xi$ and $\theta$ by the dynamics, $\varphi(\tau)=\varphi\left(\xi(\tau),\theta(\tau)\right)$.
On the other hand, by (\ref{esse}) we get 
\begin{eqnarray}
\label{pix}
p_{\xi} &=& \frac{1}{\sqrt{\xi^2 -1}}\sqrt{E^2_{\infty}
 - \frac{k^2}{\xi^2-1}} \\
\label{pth}
p_{\theta} &=&k \, ;
\end{eqnarray}
to verify the compatibility of these expressions with (\ref{sethm})
we use the equations of motion (\ref{ham}) which provide\footnote{Indeed, apart from 
the bounces against the potential walls, (\ref{ham}) describe the whole 
evolution of the system.}
\begin{equation}
\label{eqmo}
\frac{d\xi}{d\phi}= - \frac{\xi^2}{\xi^2-1}{\rm ctg}\varphi \Rightarrow 
\sqrt{\xi^2-1}\sin\varphi =c \, , 
\end{equation}
and $c$ is a constant of integration.\\
The required compatibility comes from the identification $k=E_{\infty}c$.
Since 
\begin{equation}
\label{lim}
\mathop {\lim }\limits_{\scriptstyle U \to U_{\infty}  \hfill \atop}
E_{\infty} = \left\{ 
\begin{array}{ll}
E \qquad \, \, \, \forall  \left\{ \xi, \theta \right\} \in \Gamma_H \\
i \infty \qquad  \forall \left\{ \xi, \theta \right\} \not\in \Gamma_H 
\end{array} \right.
\end{equation} 
we see by (\ref{esse}) that the solution $\varphi\left(\xi, \theta\right)$ vanishes, 
as due in presence of infinite potential walls, outside $\Gamma_H$. \\
The substitution in (\ref{2eq}) of the positions 
(\ref{sethm}) leads to the new equation
\begin{equation}
\label{cont3}
\sqrt{\xi^2-1}\cos\phi \frac{\partial r}{\partial \xi}+
\frac{\sin\phi}{\sqrt{\xi^2-1}}\frac{\partial r}{\partial \theta}=0 \, .
\end{equation}
We emphasize how this equation coincides with (\ref{cont2}), 
provided the identification $r\equiv\varrho_{\infty}$; it is just 
this correspondence between the statistical and the 
semiclassical quantum analysis to ensure that the quantum 
chaos of the Bianchi IX model approaches its deterministic one
in the considered limit. \\
Any constant function is a solution of (\ref{cont3}), but the 
normalization condition requires $r=1/4\pi$ and therefore 
we finally get 
\begin{equation}
\label{lim2}
\mathop {\lim }\limits_{\scriptstyle n \to \infty  \hfill \atop 
\scriptstyle \hbar  \to 0 \hfill}
\mid \varphi_n \mid^2  = \mid \varphi \mid^2 \equiv \varrho_{\infty} =
\left\{ \begin{array}{lll} \displaystyle
\frac{1}{4\pi} \, &\forall& \left\{ \xi, \theta\right\} \in \Gamma_H  \\ 
0  \, &\forall& \left\{ \xi, \theta\right\} \not\in \Gamma_H 
\end{array} 
\right. \, ,
\end{equation} 
say the limit for the quantum probability distribution as $n\rightarrow \infty$
and $\hbar \rightarrow 0$ associated to the wave function 
\bea
\label{swf}
\psi\left(\tau, \theta, \xi\right) &=& \varphi \left(\xi, \theta\right) e^{-i \frac{E}{\hbar}} 
=\nonumber \\
&=&\sqrt{r} ~{\rm exp} \left\{i  \int \left(p_{\xi}d\xi + p_{\theta}d\theta - E_{\infty} d\tau \right) \right\}
\eea
coincides with the classical statistical distribution on the microcanonical ensemble. \\
Though this formalism of correspondence remains valid for all Bianchi models,
only the types VIII and IX admit a normalizable wave function $\varphi(\xi, \theta)$, 
being confined in $\Gamma_H$, and a continuity equation (\ref{cont2}) which 
has a real statistical meaning. \\
Since referred to stationary states $\varphi_n(\xi, \theta)$, the considered 
semiclassical limit has to be intended in view of a ``macroscopic'' one and is not
related to the temporal evolution of the model \cite{KM97a}.

\vspace{3cm} 

\section*{Dedication}
This paper is dedicated to the memory 
of Mario Imponente (\dag 19 October 2001).

\vspace{3cm} 
-----------------------\\

The following references provide a sample of the fundamental
literature on the subject presented in this paper and 
are grouped by year of publication.

\end{document}